\documentclass[11pt,a4paper]{article}
\pdfoutput=1

\usepackage[utf8]{inputenc}

\usepackage{jheppub}

\usepackage{graphicx,amssymb,amsmath,mathtools,bm,bbm}
\usepackage[labelformat=simple]{subcaption}
\usepackage{autobreak}
\usepackage{hepunits}
\usepackage{physics}
\usepackage{epsfig}

\usepackage{hyperref}
\usepackage{url}
\hypersetup{pdfnewwindow=true}
\usepackage[utf8]{inputenc}

\def\be{\begin{equation}}
\def\ee{\end{equation}}

\allowdisplaybreaks

\preprint{
  \begin{flushright}
    MPP-2023-280 \\
    CERN-TH-2023-236
  \end{flushright}}

\title{The Higgs-gluon form factor at three loops in QCD with three mass scales}

\author[a]{Marco Niggetiedt}
\author[b]{and Johann Usovitsch}

\emailAdd{marco.niggetiedt@mpp.mpg.de}
\emailAdd{johann.usovitsch@cern.ch}
\affiliation[a]{Max-Planck-Institut f\"ur Physik, Boltzmannstra{\ss}e 8, 85748 Garching, Germany}
\affiliation[b]{Theoretical Physics Department, CERN, 1211 Geneva, Switzerland}

\abstract{We report on the complete three-loop corrections to the Higgs-gluon 
form factor in QCD. While previous calculations are based on QCD with a 
single heavy quark of arbitrary mass, we extend the study to QCD involving 
two different massive quark flavors. Thereby, the full set of possible 
Feynman diagrams at three-loop order is taken into account. Employing 
differential equations for the relevant master integrals, we determine the 
form factor in terms of analytic expansions. Outside the radii of convergence, 
we compute high-precision numerical samples over the two-dimensional physical 
parameter space. Our new findings will enter as virtual corrections 
the computation of the top-bottom interference in hadronic Higgs-boson 
production at next-to-next-to-leading order (NNLO) in QCD.}

\begin{document}

\maketitle

\newpage


\section{Introduction} \label{sect:intro}

After the discovery of the Higgs-boson at the Large Hadron Collider (LHC) in 2012 by the collaborations ATLAS and CMS~\cite{ATLAS:2012yve,CMS:2012qbp}, high energy physics has entered the phase of precision measurements of properties of the Higgs-boson. In this context, precise theory predictions for Higgs observables play an indispensable role in the search for footprints of physics beyond the Standard Model (SM). Recent experimental measurements of Higgs-boson couplings~\cite{ATLAS:2022vkf,CMS:2022dwd} with SM particles of masses ranging over multiple orders of magnitude are in agreement with the SM expectations. In view of the growth of available data as well as the high-luminosity phase of the LHC, the associated statistical uncertainties will decrease, requiring a reduction of theory uncertainties in order to keep up with the experimental precision. 

The predominant mode for the production of a Higgs-boson at the LHC is gluon fusion~\cite{LHCHiggsCrossSectionWorkingGroup:2016ypw}. Its total cross section exceeds the cross section of any other production channel by at least one order of magnitude at the typical center-of-mass energies. The effects of massive quarks beyond next-to-leading order (NLO) account for roughly one third of the total theory uncertainty imposed on the gluon fusion cross section. Being a loop-induced process, it is sensitive to physics beyond the SM. Therefore, a firm understanding of the Higgs production cross section is mandatory for current and future precision physics. 

One of the building blocks in the calculation of the total cross section is the Higgs-gluon form factor parametrizing the amplitude for the scattering of two on-shell gluons with a possibly off-shell Higgs-boson in QCD. While at the two-loop order, the form factor is known for quarks of arbitrary mass running in the loops, both in numerical and analytical form~\cite{Spira:1995rr,Harlander:2005rq,Anastasiou:2006hc,Aglietti:2006tp}, addressing the exact effects of massive quarks at three loops becomes more challenging for reasons concerning the reduction of symbolic expressions as well as the emergence of new mathematical structures of the underlying integrals. 

To overcome these difficulties, the form factor at three loops was first approximated as an expansion for heavy quarks~\cite{Harlander:2009bw,Pak:2009bx}, which is sufficient for the characterization of top quark mass effects. We note that even at four loops, the expansion of the form factor in the limit of a heavy quark mass in known~\cite{Davies:2019wmk}. In order to include the impact of the lighter quark flavors at three loops, the expansion around the large mass limit was matched with the non-analytic terms at threshold, calculated in Ref.~\cite{Grober:2017uho}, and convergence was improved using Pad\'e approximants~\cite{Davies:2019nhm}. Subsequently, the full exact result for a single massive quark flavor of arbitrary mass became available by numerically solving the relevant integrals~\cite{Czakon:2020vql}. Up to now, the corresponding analytical result has only been feasible for certain subsets of Feynman diagrams~\cite{Harlander:2019ioe,Prausa:2020psw}. 
\begin{figure}[t]
    \centering
    \begin{subfigure}[b]{.2\textwidth}
        \centering
        \includegraphics[scale=.15]{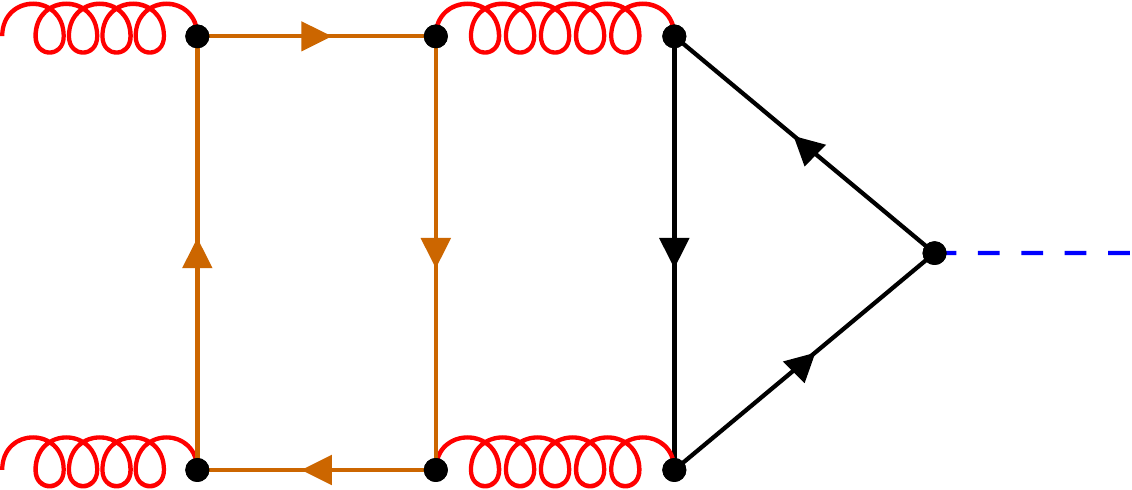}
    \end{subfigure}%
        \begin{subfigure}[b]{.2\textwidth}
        \centering
        \includegraphics[scale=.15]{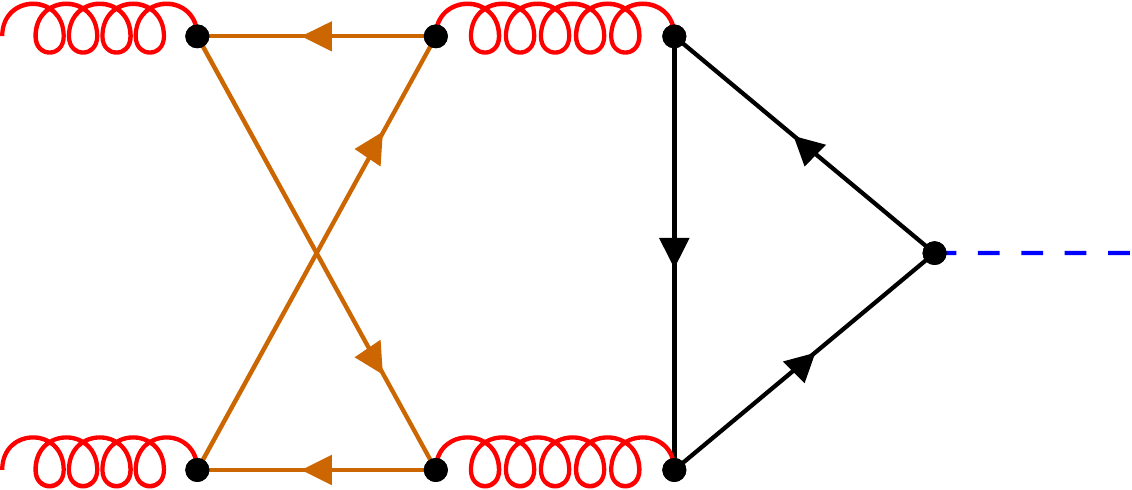}
    \end{subfigure}%
        \begin{subfigure}[b]{.2\textwidth}
        \centering
        \includegraphics[scale=.15]{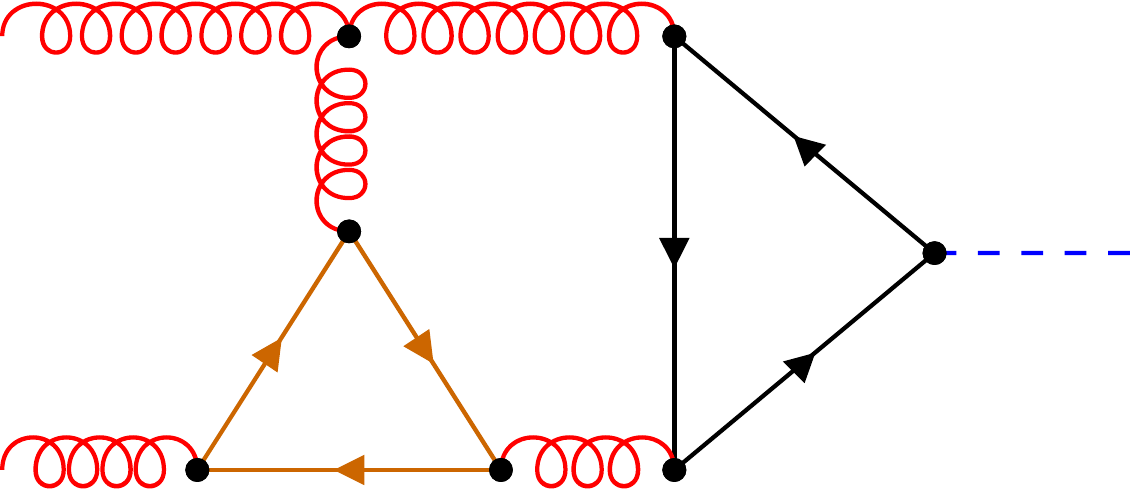}
    \end{subfigure}%
    			\begin{subfigure}[b]{.2\textwidth}
        \centering
        \includegraphics[scale=.15]{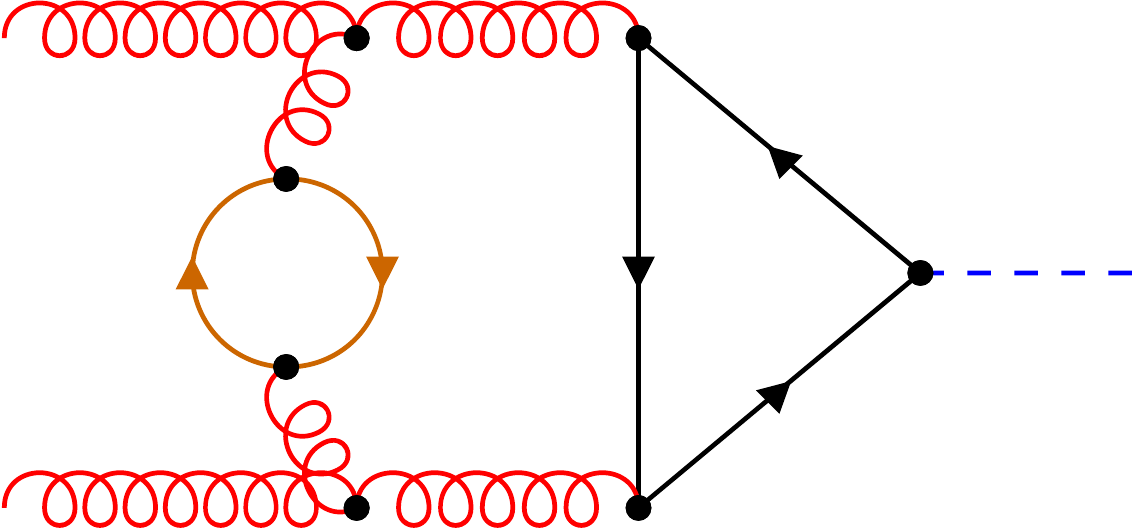}
    \end{subfigure}%
    			\begin{subfigure}[b]{.2\textwidth}
        \centering
        \includegraphics[scale=.15]{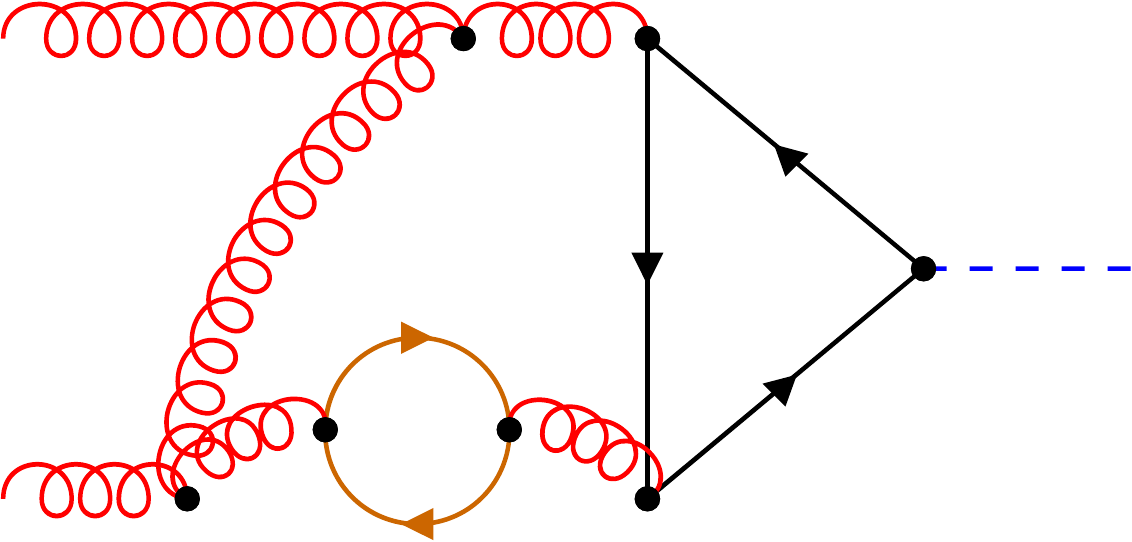}
    \end{subfigure}%
    \\[14pt]
    \centering
    \begin{subfigure}[b]{.2\textwidth}
        \centering
        \includegraphics[scale=.12]{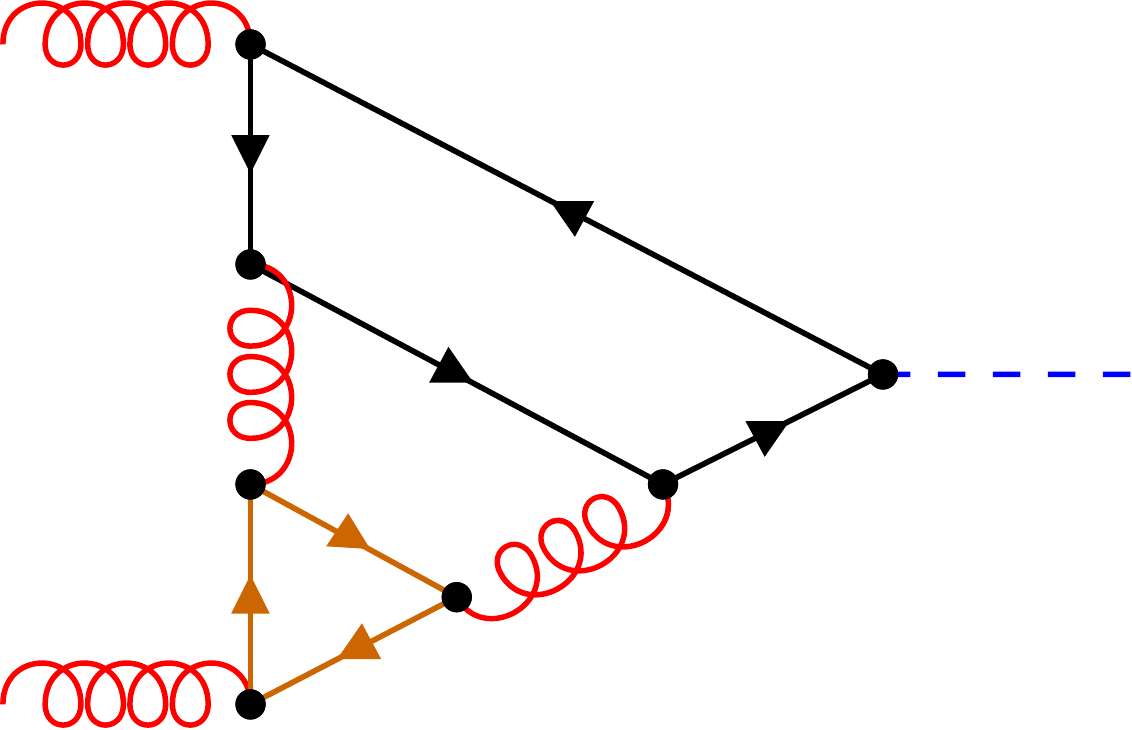}
    \end{subfigure}%
        \begin{subfigure}[b]{.2\textwidth}
        \centering
        \includegraphics[scale=.12]{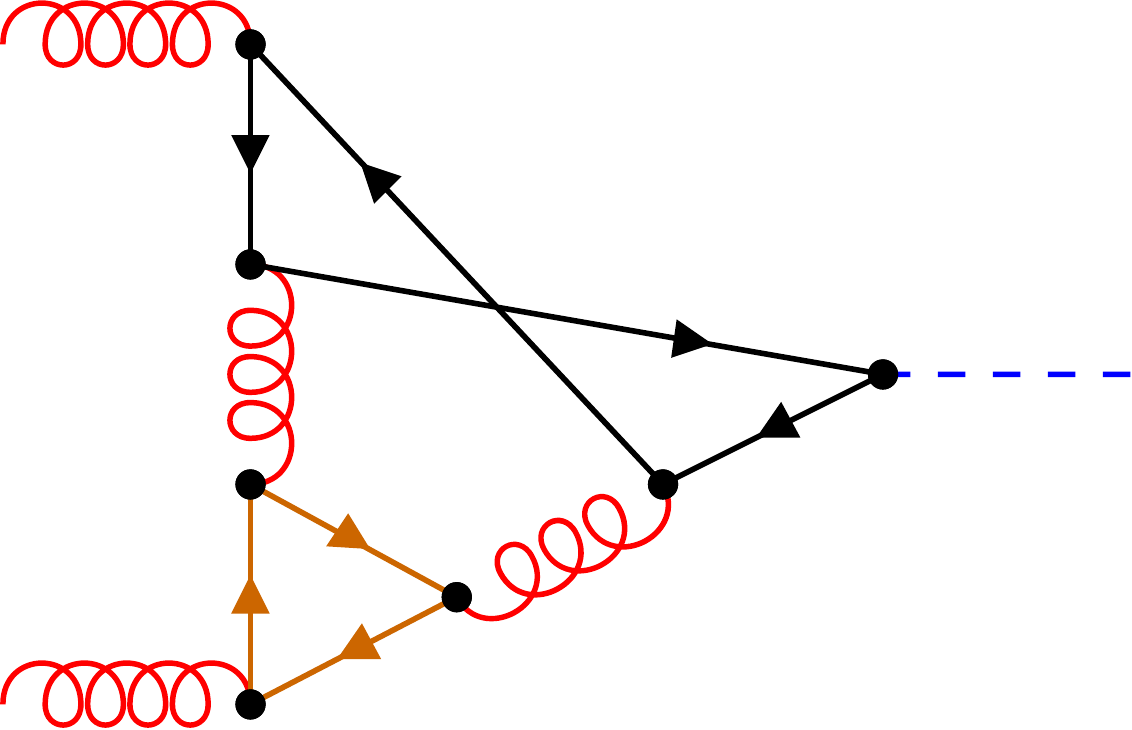}
    \end{subfigure}%
        \begin{subfigure}[b]{.2\textwidth}
        \centering
        \includegraphics[scale=.12]{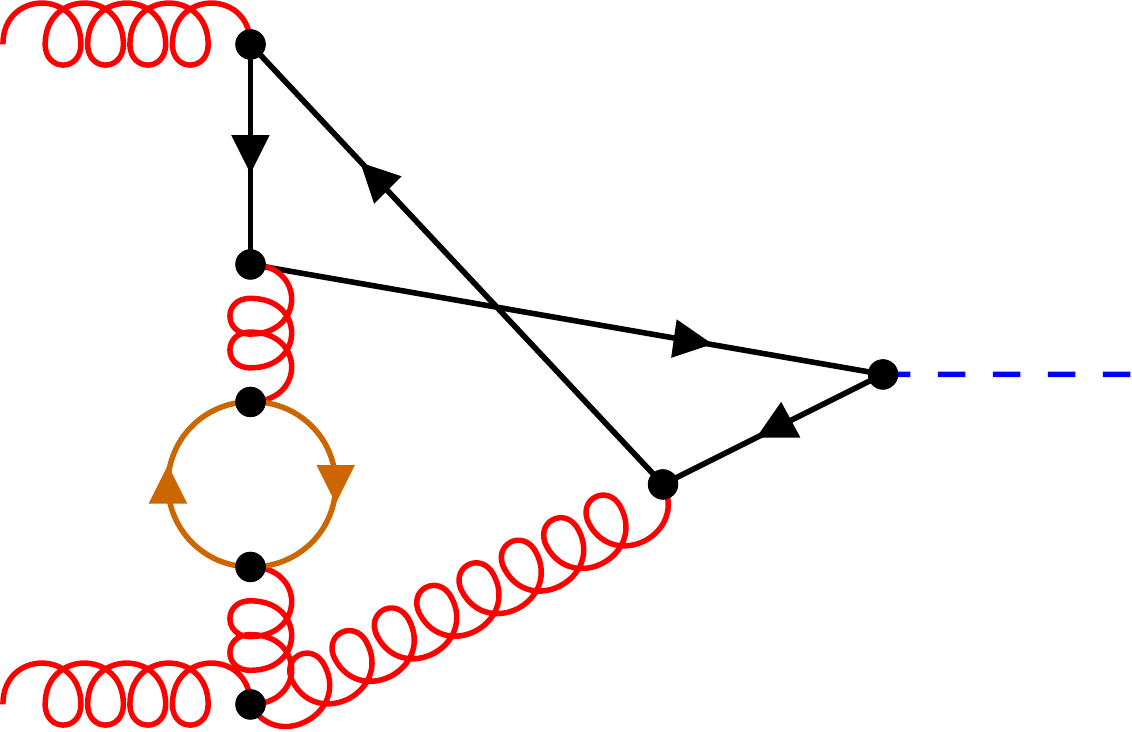}
    \end{subfigure}%
    			\begin{subfigure}[b]{.2\textwidth}
        \centering
        \includegraphics[scale=.12]{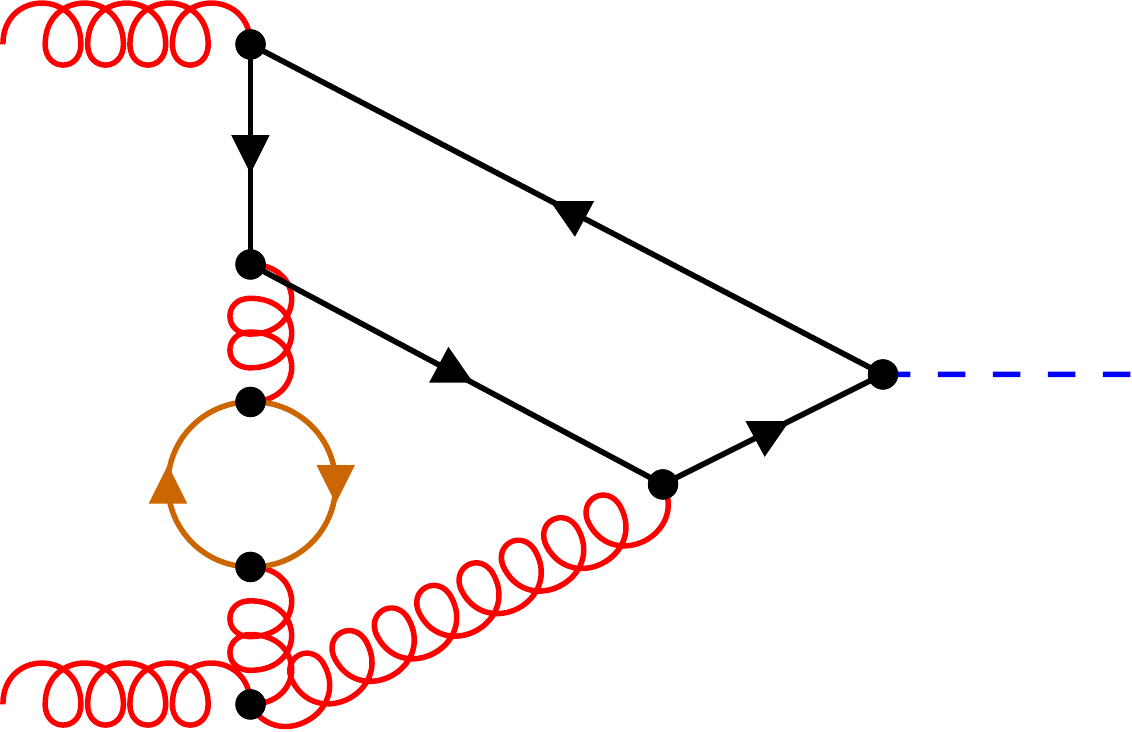}
    \end{subfigure}%
    			\begin{subfigure}[b]{.2\textwidth}
        \centering
        \includegraphics[scale=.12]{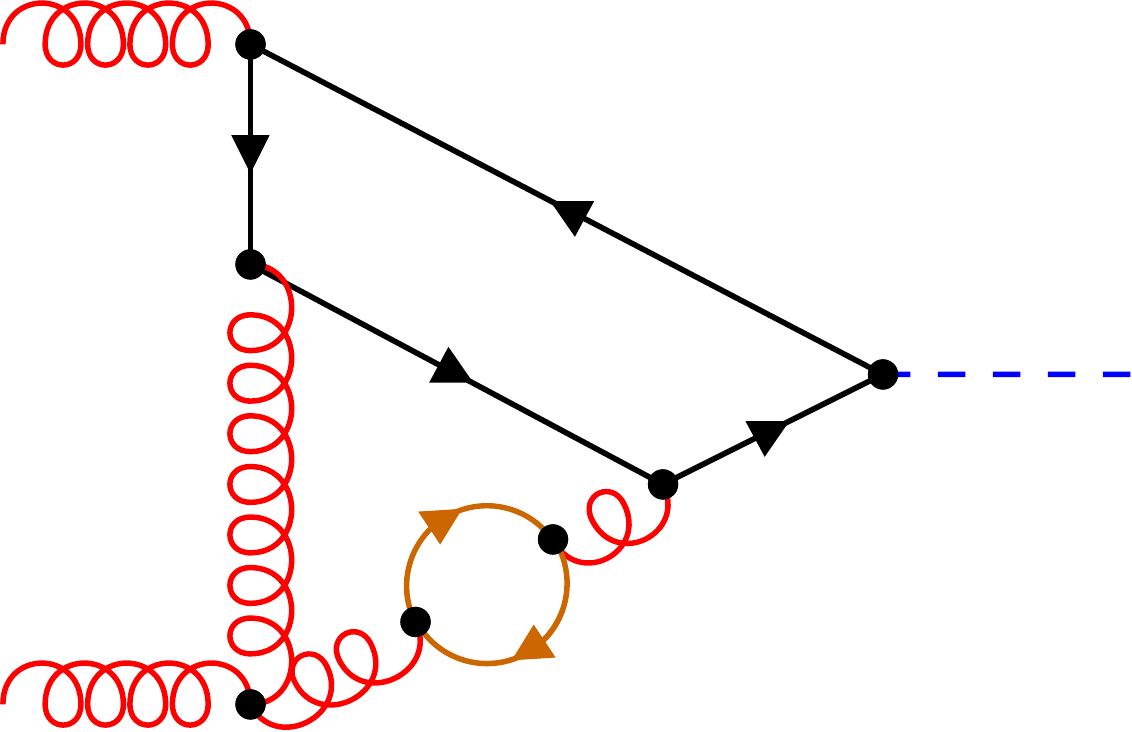}
    \end{subfigure}%
    \\[14pt]
    \centering
    \begin{subfigure}[b]{.2\textwidth}
        \centering
        \includegraphics[scale=.12]{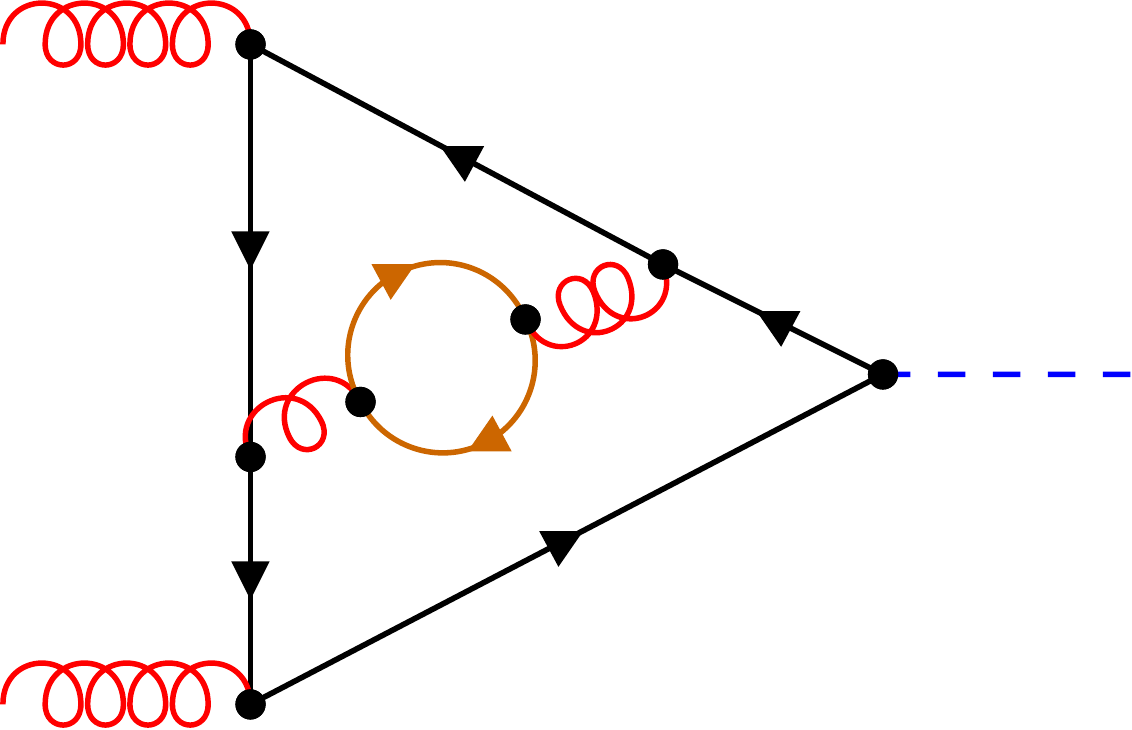}
    \end{subfigure}%
        \begin{subfigure}[b]{.2\textwidth}
        \centering
        \includegraphics[scale=.12]{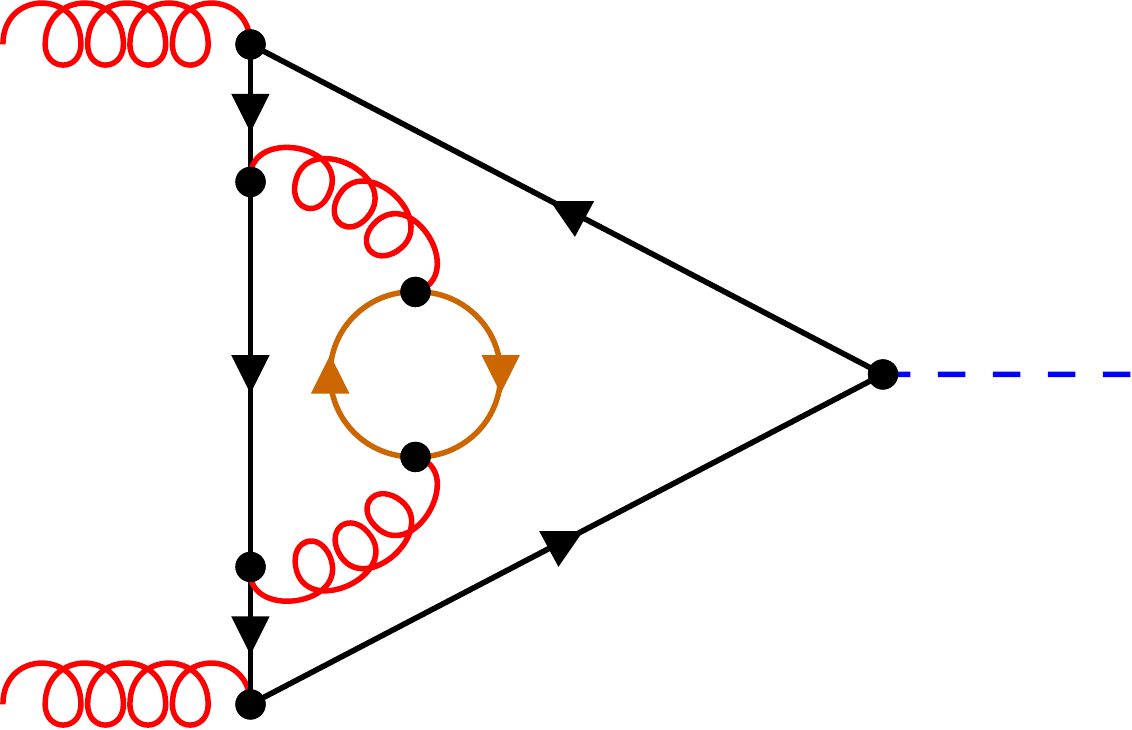}
    \end{subfigure}%
        \begin{subfigure}[b]{.2\textwidth}
        \centering
        \includegraphics[scale=.12]{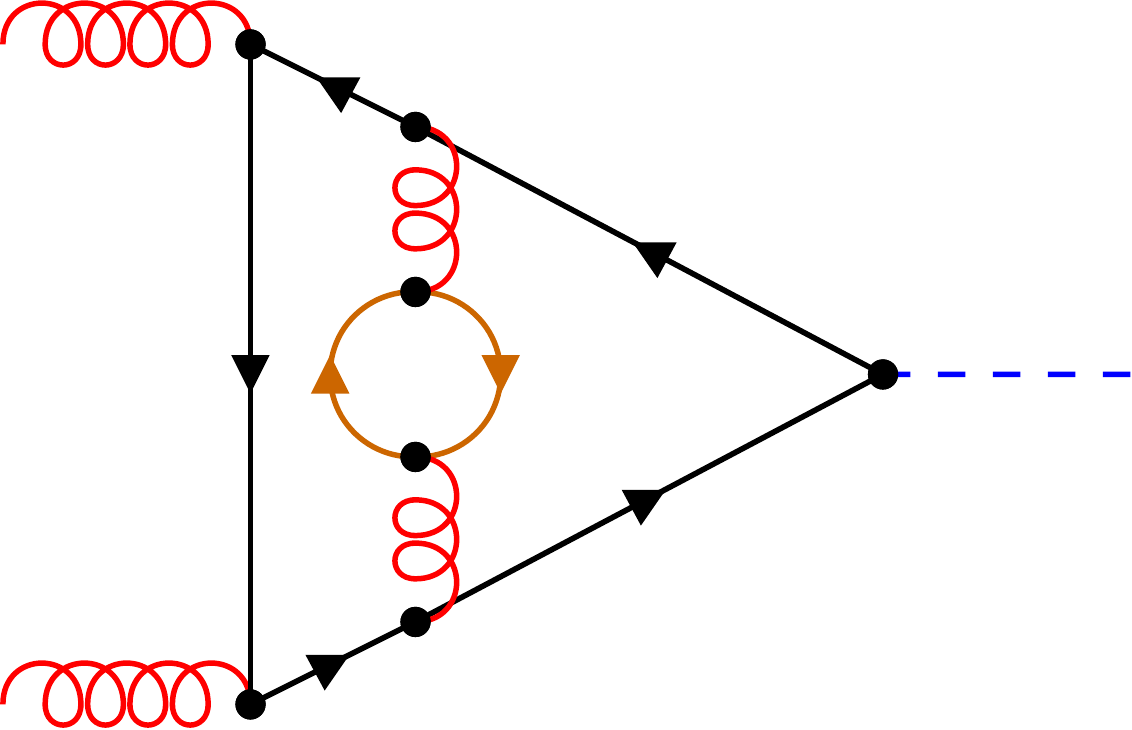}
    \end{subfigure}%
    			\begin{subfigure}[b]{.2\textwidth}
        \centering
        \includegraphics[scale=.12]{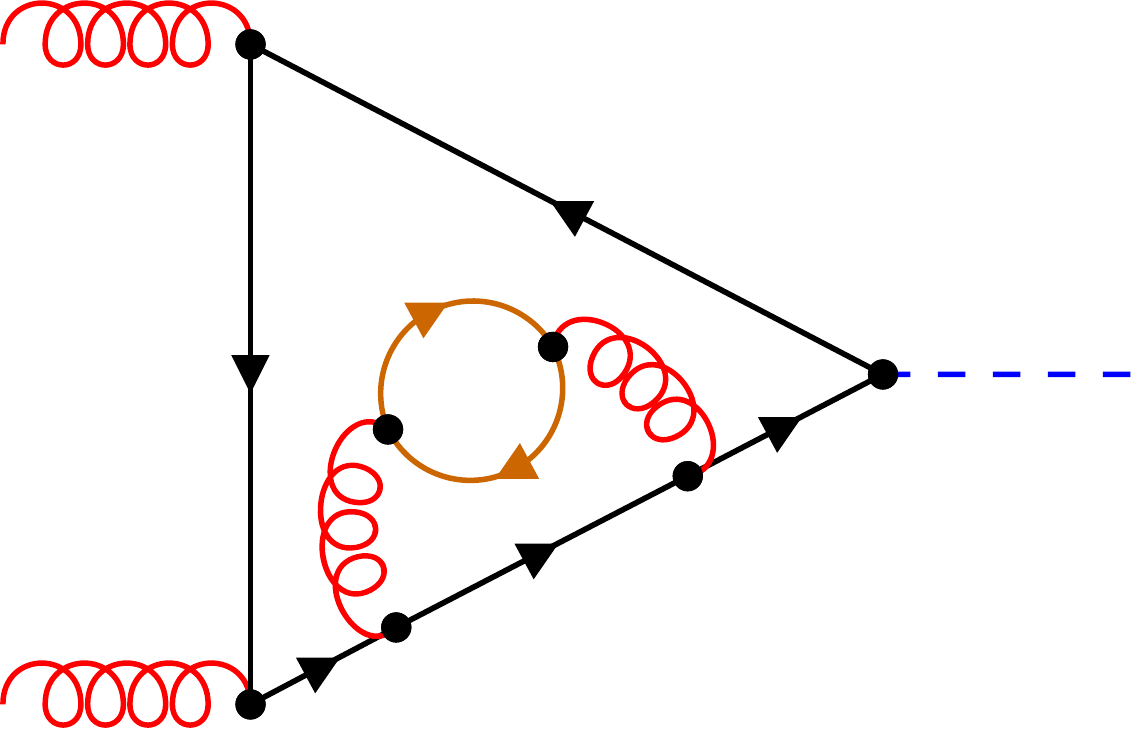}
    \end{subfigure}%
    			\begin{subfigure}[b]{.2\textwidth}
        \centering
        \includegraphics[scale=.15]{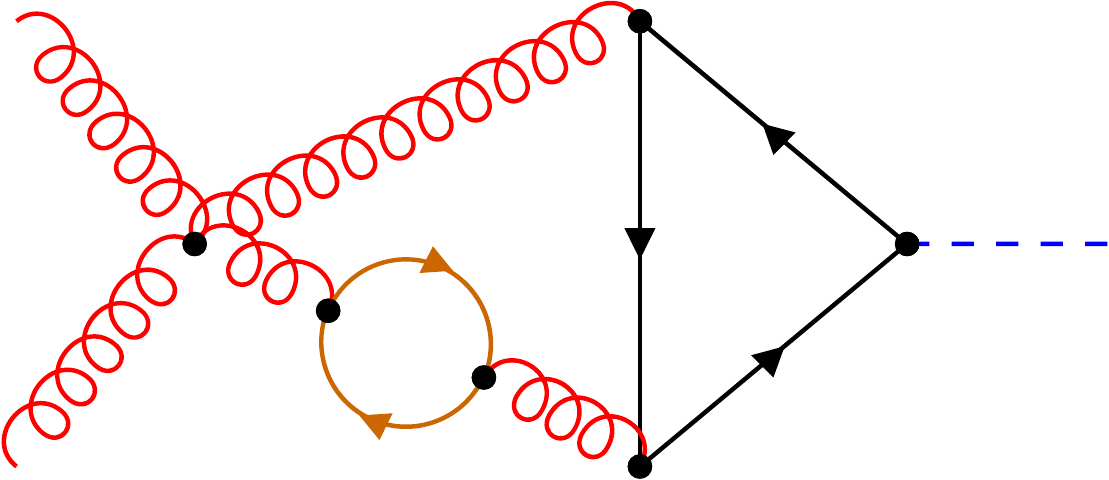}
    \end{subfigure}%
\caption{The complete set of three-loop Feynman diagrams relevant to the Higgs-gluon form factor featuring two quark loops. While the quark coupling to the external Higgs-boson must be massive, the quark running in the other loop can either be massive or massless. In the present work, we focus on the general case involving two heavy quark flavors with possibly different masses in the loops.}
\label{fig:intro:diags}
\end{figure}

In the paper at hand, we are concerned with a new class of Feynman diagrams entering the Higgs-gluon form factor at three loops in QCD. It is important to emphasize that the exact calculation in Ref.~\cite{Czakon:2020vql} takes into account the effect of a single massive quark flavor. However, starting from three loops, Feynman diagrams can develop more than just one closed fermion loop. Hence, in order to completely describe the form factor at three loops with general quark masses, a treatment of the subset of Feynman diagrams incorporating two different massive quark flavors as shown in Fig.~\ref{fig:intro:diags} is obligatory. In the following we present their computation consisting of a combination of numerical and analytical methods, both of which employ differential equations for the relevant integrals. The paper is structured as follows. We introduce our notation and definitions relevant for the understanding in Sect.~\ref{sect:definition}. Subsequently, the methods used throughout the calculation are reviewed in Sect.~\ref{sect:calc}. Our results are discussed in Sect.~\ref{sect:results} and we draw conclusions in Sect.~\ref{sect:conclusion}. 

\section{Notation and definitions} \label{sect:definition}

We consider corrections to the Higgs-gluon form factor parametrizing the amplitude for the production of a Higgs-boson in the scattering of two on-shell gluons carrying momenta $p_{1,2}$, helicities $\lambda_{1,2}$, and colors $a_{1,2}$. Since the present work is intended to conclude the study of the form factor performed in Ref.~\cite{Czakon:2020vql}, we closely follow the definitions thereof and adopt the notation to a large extent. Hence, we define the bare form factor $C$ accordingly: 
\begin{multline} \label{eq:definition:amplitude}
-i \mathcal{M}\big[ g(p_1,\lambda_1,a_1) + g(p_2,\lambda_2,a_2) \to H
\big] \equiv \\[.2cm]
i \delta^{a_1a_2} \big[ (\epsilon_1 \cdot p_2) \, (\epsilon_2
\cdot p_1) - (\epsilon_1 \cdot \epsilon_2)  \, (p_2\cdot p_1) \big] \,
\frac{1}{v} \frac{\hat{\alpha}_s}{\pi} \, C \,.
\end{multline}
Since in the SM, the amplitude under consideration is loop-induced, an overall factor of the bare strong coupling constant $\hat{\alpha}_s$ has been factored out. Likewise, the vacuum expectation value, $v$, originating from the coupling of the Higgs-boson to massive quarks is isolated. The gluon polarization vectors are normalized in the usual way. 

The kinematic configuration of the underlying scattering process gives rise to a single kinematic variable $s=2 \, p_1 \cdot p_2$ coinciding with the virtuality of the potentially off-shell Higgs-boson, which is in the following denoted by $m_H^2$. The unrenormalized form factor $C$ admits a perturbative expansion in the bare strong coupling constant: 
\begin{equation}
C = \sum_{n=0}^{\infty} \left(\frac{\hat{\alpha}_s}{\pi}\right)^n C^{(n)} \,.
\end{equation}
Owing to the fact that massive particles are required to mediate the interaction between the massless initial-state gluons and the Higgs-boson, the Higgs-gluon form factor naturally becomes a function of mass parameters. In the paper at hand, we focus on to the calculation of the predominant QCD corrections at three loops retaining the mass of two different quark flavors. In particular, our analysis rests on QCD with $n_f=n_t+n_b+n_l$ quark flavors and we refer to the massive flavors as the top quark and the bottom quark, respectively, while the remaining quarks are treated as massless particles. However, in the subsequent discussion, the masses of the top and bottom quark should be understood as variable parameters. In order to keep track of contributions to the form factor stemming from different classes of Feynman diagrams, we refrain from setting the labels $n_t$ and $n_b$ to their physical values. 

With the aid of the introduced flavor-labels indicating the number of top ($n_t$), bottom ($n_b$) and massless ($n_l$) quark loops in the underlying Feynman diagrams, we decompose the three-loop coefficient of the form factor accordingly: 
\begin{equation} \label{eq:definition:decomposition}
\begin{split}
C^{(2)} = \; & n_t \, C^{(2)}_{t} + n_t^2 \, C^{(2)}_{tt} + n_t n_l \, C^{(2)}_{tl} \\
        + \, & n_b \, C^{(2)}_{b} + n_b^2 \, C^{(2)}_{bb} + n_b n_l \, C^{(2)}_{bl} \\
        + \, & n_t n_b \left(C^{(2)}_{tb} + C^{(2)}_{bt}\right) \,.
\end{split}
\end{equation}
Contributions originating from Feynman diagrams with a single closed top or bottom quark loop are captured with the coefficients $C^{(2)}_{t}$ and $C^{(2)}_{b}$, respectively. As opposed to the corrections to the form factor at lower orders in perturbation theory, the three-loop corrections develop for the first time contributions of the type depicted in Fig.~\ref{fig:intro:diags}, namely, Feynman diagrams with two closed fermion loops. In QCD with a single massive quark flavor, the only possible configurations are given by either assigning the massive flavor to both fermion loops, or allowing for a massless flavor in the second fermion loop while the Higgs must couple to the massive quark. In the decomposition of the form factor in Eq.~\eqref{eq:definition:decomposition} valid for two massive quark flavors, the former contribution is covered by the coefficients $C^{(2)}_{tt}$ and $C^{(2)}_{bb}$, while the latter one is accounted for in terms of $C^{(2)}_{tl}$ and $C^{(2)}_{bl}$, respectively. Here, the first letter in the subscript refers to the quark flavor attached to the Higgs-boson. 

We point out that the subset of Feynman diagrams featuring one massless fermion loop in addition to the massive quark loop has been addressed in Ref.~\cite{Harlander:2019ioe} by means of analytical techniques, such that $C^{(2)}_{tl}$ and $C^{(2)}_{bl}$ are known in terms of harmonic polylogarithms. Furthermore, the complete set of three-loop Feynman diagrams in QCD with a single massive quark flavor has been subject to the exact computation in Ref.~\cite{Czakon:2020vql} based on semi-analytical methods. Hence, all coefficients that appear in the first and second line on the right-hand side of Eq.~\eqref{eq:definition:decomposition} are known either in terms of analytic functions or expansions supplemented with high-precision numerical samples over a one-dimensional parameter space. 

However, it is important to stress that a complete knowledge of the three-loop QCD corrections, $C^{(2)}$, in the most general case requires the calculation of terms in the third line of Eq.~\eqref{eq:definition:decomposition} in addition to the aforementioned constituents. The missing contribution arises in QCD with two massive quark flavors. More precisely, a class of Feynman diagrams with two distinct massive quarks running in the loops as illustrated in Fig.~\ref{fig:intro:diags} emerges from the underlying theory. Thus, the new contribution being proportional to $n_t n_b$ consists of two building blocks denoted by the coefficients $C^{(2)}_{tb}$ and $C^{(2)}_{bt}$, where in analogy to the other coefficients, the first letter in the subscript indicates the quark flavor that couples directly to the Higgs-boson, whereas the second letter refers to the quark flavor running in the second loop. The calculation of these two coefficients is at the core of the present paper. 

First, we note that in contrast to the known coefficients in Eq.~\eqref{eq:definition:decomposition} capturing Feynman diagrams with only one massive quark flavor, the new contributions, $C^{(2)}_{tb}$ and $C^{(2)}_{bt}$, require the introduction of an additional dimensionless scale. Consequently, the Higgs-gluon form factor becomes a function of two dimensionless variables, which we define as the individual quark masses rescaled with respect to the mass of the Higgs-boson squared: 
\begin{equation} \label{eq:definition:variables}
x = \frac{m_t^2}{m_H^2}
\,, \qquad
y = \frac{m_b^2}{m_H^2} \,.
\end{equation}
Inspired by the fact that three mass scales enter the parametrization, we dub the coefficients under consideration, $C^{(2)}_{tb}$ and $C^{(2)}_{bt}$, three-scale coefficients in the remainder of the text. The three-scale coefficients depend on $x$ and $y$ as arguments: 
\begin{equation}
C^{(2)}_{tb} \equiv C^{(2)}_{tb}(x,y)
\,, \qquad
C^{(2)}_{bt} \equiv C^{(2)}_{bt}(x,y) \,.
\end{equation}
It is obvious, that the set of underlying Feynman diagrams contributing to $C^{(2)}_{bt}$ is a copy of diagrams accounted for in terms of $C^{(2)}_{tb}$ with flavors of the quarks swapped. Therefore, the three-scale coefficients are related by symmetry: 
\begin{equation} \label{eq:definition:sym}
C^{(2)}_{tb}(x,y) = C^{(2)}_{bt}(y,x) \,.
\end{equation}
In our calculation, we profit from the fact that knowing one of the two coefficients in the entire parameter space spanned by $x$ and $y$ is sufficient for a conclusive determinations of the Higgs-gluon form factor at three loops. In accordance with Eq.~\eqref{eq:definition:sym}, the coefficients are related to each other by simultaneously exchanging flavors and variables. 

In order to yield a physically meaningful result, the bare form factor must be renormalized by redefinition of couplings and masses. The strong coupling constant is renormalized in the $\overline{\mathrm{MS}}$ scheme with $n_f=n_t+n_b+n_l$ flavors. Additionally, we perform the decoupling of the massive quark flavors using the decoupling constant derived in Refs.~\cite{Bernreuther:1981sg,Larin:1994va}. Consequently, the strong coupling constant $\alpha_s \equiv \alpha_s^{(n_l)}(\mu)$ exhibits a dependence on the renormalization scale $\mu$ dictated by the $\beta$-function for $n_l$ massless quarks. The gluon wave function as well as the masses of the heavy quarks are renormalized on-shell. To this end, we recomputed the relevant renormalization constants up to two loops to sufficiently high powers in the dimensional regulator $\varepsilon$. In principle, the on-shell gluon wave function renormalization constant can be taken from Ref.~\cite{Czakon:2007wk} since at two loops, the contribution originating from a second massive flavor adds up linearly to the result with one heavy flavor. Likewise, $Z_m$ has been computed in the on-shell scheme in Ref.~\cite{Gray:1990yh}, however, in the presence of a second heavy quark flavor only up to the finite part in the dimensional regulator. The integrals needed for the construction of a mass-counterterm in the present calculation have been solved in Ref.~\cite{Davydychev:1998si} in closed form in $\varepsilon$ and the masses. We point out, that the corrections to the Higgs-gluon form factor at one and two loops are provided analytically to sufficiently high orders in $\varepsilon$ in Ref.~\cite{Anastasiou:2020qzk}, such that the derivation of the UV-counterterm is straightforward. 

After UV-renormalization, the form factor still exhibits IR divergences regularized in terms of poles in $\varepsilon$. The IR singularities can be factorized employing the $I$-operator of Ref.~\cite{Catani:1998bh}, which is explicitly given for the form factor at hand in Ref.~\cite{deFlorian:2012za}. Alternatively, the IR singularities may be isolated in the $\overline{\mathrm{MS}}$ factorization scheme \cite{Becher:2009cu,Becher:2009qa}, such that only the poles in $\varepsilon$ are absorbed. In Ref.~\cite{Czakon:2020vql}, the three-loop corrections to the Higgs-gluon form factor were presented in terms of a finite remainder defined with the aid of the $I$-operator and the translation to the $\overline{\mathrm{MS}}$ factorization scheme was provided for convenience. 

We point out that the finite remainder for the three-scale coefficients considered in this paper is independent of the particular scheme choice. Therefore, we define finite remainders, $\mathcal{C}^{(2)}_{tb}$ and $\mathcal{C}^{(2)}_{bt}$, without specifying the scheme choice and implicitly treat the form factor in the $\overline{\mathrm{MS}}$ factorization scheme. Moreover, it is worth to mention that the finite remainders of the corresponding three-scale coefficients are devoid of explicit logarithms of the renormalization scale. 

\section{Calculation} \label{sect:calc}

The computation of the three-scale coefficients is divided into multiple steps aligned with the workflow of current multi-loop calculations. First, all possible three-loop Feynman diagrams for our process of interest featuring two different heavy quark loops are generated within the framework of \texttt{DiaGen/IdSolver}~\cite{DiaGen}. In doing so, not only the relevant diagrams are generated, but also a graph-theoretical comparison based on the topologies of the corresponding diagrams is performed to allow for the construction of a minimal set of integral families. In parallel, computer code for the matching of scalar expressions with the families of integrals after projection to the form factor is automatically produced. The projection of the sum of Feynman diagrams to the form factor consists of working out the Lorentz and color algebra, which is performed with \texttt{FORM}~\cite{Vermaseren:2000nd,Kuipers:2012rf,Ruijl:2017dtg} and the package \texttt{color}~\cite{vanRitbergen:1998pn}, and the matching with the previously defined integral families. Thereby, a linear combination of scalar integrals with coefficients being rational functions in $\varepsilon$, $x$ and $y$ as defined in Eq.~\eqref{eq:definition:variables} is obtained. Employing Integration-By-Parts identities~\cite{Tkachov:1981wb,Chetyrkin:1981qh} and the algorithm of Laporta~\cite{Laporta:2000dsw}, the set of scalar integrals can be reduced to a set of 290 master integrals. We point out that this set comprises all master integrals entering both three-scale coefficients. The reduction to master integrals can efficiently be accomplished with \texttt{Kira}~\cite{Maierhofer:2017gsa,Maierhofer:2018gpa,Klappert:2020nbg} in combination with \texttt{FireFly}~\cite{Klappert:2019emp,Klappert:2020aqs}. We decide to choose a basis of master integrals, such that in all expressions, the dependence on the dimensional regulator factorizes from the dependence on the kinematic variables and masses in denominators of coefficients in front of these master integrals~\cite{Smirnov:2020quc,Usovitsch:2020jrk}. 

At the heart of the evaluation of the three-scale coefficients lies the efficient computation of the master integrals. In order to tackle the computation thereof, we devise two complementary strategies, one of which is based on deep asymptotic expansions of the relevant integrals and the other one rests on numerical techniques as pursuing a fully analytical approach is out of reach with current technology. Both strategies rely on the method of differential equations~\cite{Kotikov:1990kg,Kotikov:1991pm,Kotikov:1991hm,Remiddi:1997ny}. We note that in order to set up closed systems of first-order linear differential equations in $x$ and $y$ for the master integrals appearing in the form factor, the basis of master integrals must be extended to 305 integrals. In analogy to the strategy outlined in Ref.~\cite{Czakon:2020vql}, we insert truncated $\varepsilon$-expansions for the master integrals into the differential equations and solve for the expansions coefficients, which no longer depend on the dimensional regulator. The truncation order is determined by poles in $\varepsilon$ in the form factor and the differential equations, such that the finite part of the three-scale coefficients can be obtained. 

The first strategy for solving the master integrals is based on asymptotic expansions and exploits the fact that for physical masses of the heavy quarks and the Higgs-boson, respectively, a hierarchy among the scales is implied. In particular, we consider the full set of master integrals and impose a hierarchy in accordance with $m_b^2 \ll m_H^2 \ll m_t^2$. Taking advantage of the hierarchy allows us to subsequently expand the master integrals in a diagrammatic language for large $m_t^2$ and the resulting expansion coefficients in small $m_b^2$ using their parametric representation. By successively expanding the master integrals in the double-limit, the dependence on the different scales is factorized, such that only single-scale integrals must be computed. The motivation behind this approach is multi-fold. On the one hand, it significantly reduces the complexity of the calculation in comparison to a fully analytic procedure, which might not even be feasible. On the other hand, there is no need to meticulously consider the analytic continuation of analytic expressions or to rationalize roots, which is usually necessary when internal propagators carry masses. Moreover, the asymptotic expansion attains the form of a power-logarithmic series, such that an efficient evaluation is guaranteed. 

In the first step, we apply the asymptotic large mass expansions to the set of master integrals. The corresponding diagrammatic formulation~\cite{Gorishnii:1989dd,Smirnov:1990rz,Smirnov:1994tg,Smirnov:2002pj} for the asymptotic estimate of an integral $F$ with graphical representation $\Gamma$ in the limit $m_t^2 \to \infty$ reads 
\begin{equation} \label{eq:calc:LME}
    F_\Gamma(\{p_j\}_\Gamma, m_b^2, m_t^2)
    \sim
    \sum\limits_\gamma
    F_{\Gamma/\{\gamma_1 \cup \ldots \cup \gamma_n\}}(\{p_j\}_\Gamma, m_b^2)
    \circ
    \prod\limits_i
    \mathcal{T}_{\gamma_i}
    F_{\gamma_i}(\{p_j\}_\gamma, m_b^2, m_t^2) \,,
\end{equation}
where $\gamma_i$ denotes an asymptotically irreducible subgraph, $\{p_j\}_\gamma$ is the set of momenta being external with respect to the subgraph, and $\mathcal{T}_{\gamma_i}$ refers to a Taylor operator, which expands in all quantities considered small in this context. In this way, the original three-scale integral factorizes into a single-scale vacuum integral and a reduced vertex-function, which is independent of the heavy scale $m_t^2$. The identification of subgraphs has been implemented in the framework of \texttt{DiaGen/IdSolver}. After application of the Taylor operators on the right-hand side of Eq.~\eqref{eq:calc:LME}, the expressions can be organized in terms of scalar integrals rendering possible a reduction to master integrals. 

It turns out that all master integrals we encounter in the large mass expansions were previously studied in the literature. The vacuum integrals can be at most three-loop integrals, which have been computed in Ref.~\cite{Schroder:2005va}, whereas the most complicated reduced vertex-functions are two-loop integrals investigated analytically, for example, in Refs.~\cite{Anastasiou:2006hc,Aglietti:2004tq,vonManteuffel:2017hms}. Although the relevant two-loop integrals are, in principle, available in analytic form, we point out that the two crossed vertex-integrals studied in Ref.~\cite{vonManteuffel:2017hms} are expressed in terms of iterated integrals over elliptic kernels. We determined all two-scale integrals by means of differential equations in terms of deep expansions with numerical coefficients in special kinematic points and sampled them over the physical parameter space. 

For the sake of presentation of the form factor, we find it more beneficial to further exploit the hierarchy of scales and to analytically expand the vertex-functions that emerge from the asymptotic large mass expansion in the limit of small internal mass. To this end, we employ a variety of different techniques. For most of the integrals, we make use of the method of regions~\cite{Beneke:1997zp,Smirnov:1998vk,Smirnov:1999bza} in accordance with its geometric formulation in Feynman parameter space~\cite{Pak:2010pt,Jantzen:2012mw}. The resulting integrals in different scaling regions are evaluated with the aid of \texttt{HyperInt}~\cite{Panzer:2014caa}. Alternatively, the integrals can be treated numerically with Mellin-Barnes methods implemented in the package \texttt{MB.m}~\cite{Czakon:2005rk}, where the analytic results are restored via the PSLQ algorithm~\cite{PSLQ}. 

However, the sketched method only provides the first few expansion coefficients in the double-limit of large $m_t^2$ and small $m_b^2$. We would like to extend the depth of the series expansion to high order, rendering possible a precise description of the form factor within a certain radius of convergence. In particular, we seek for a deep asymptotic expansion of the bare three-scale coefficient at $x\to\infty$, 
\begin{equation} \label{eq:calc:exp}
C^{(2)}_{tb}(x,y) = \sum_{n=n_0}^{\overline{n}} \sum_{m=0}^{\overline{m}} c_{nm}(y) \frac{\log^m(x)}{x^n} \,,
\end{equation}
with truncation order $\overline{n}$ and maximum logarithmic power $\overline{m}$. Here, the $y$-dependent expansion coefficients are understood as asymptotic expansions in $y\to 0$ themselves, 
\begin{equation} \label{eq:calc:coeff}
c_{nm}(y) = \sum_{k=k_0}^{\overline{k}} \sum_{l=0}^{\overline{l}} c^{(nm)}_{kl} y^k \log^l(y) \,,
\end{equation}
truncated at $\overline{k}$ and maximum logarithmic power $\overline{l}$. The other three-scale coefficient, $C^{(2)}_{bt}$, is treated analogously. The individual master integrals admit asymptotic expansions similar to Eq.~\eqref{eq:calc:exp}. In order to determine the expansion coefficients of the master integrals and in turn the coefficients $c_{nm}(y)$, we exploit the fact that a system of first-order linear differential equations for the set of master integrals provides relations among the coefficients $c_{nm}(y)$ at different values of $n$ and $m$. Hence, the expansion coefficients can be expressed in terms of a finite set of boundary coefficients, which can be inferred from the leading asymptotic behavior as introduced in Eq.~\eqref{eq:calc:LME}. Instead of determining the $y$-dependent expansion coefficients at the level of master integrals, we found it more advantageous to work out the expansion of the three-scale coefficients multiple times at different rational values for $y$ and reconstruct their analytic form at the level of the form factor coefficients. In practice, we truncate the expansions of $C^{(2)}_{tb}$ and $C^{(2)}_{bt}$, respectively, at $\overline{n}=40$ and $\overline{k}=50$, which is more than sufficient for the phenomenological application considering $m_t$ to take the physical value of the top quark mass and $m_b$ to assume the mass of a lighter quark flavor. 
\begin{figure}[t]
\center
\includegraphics[width=0.45\textwidth]{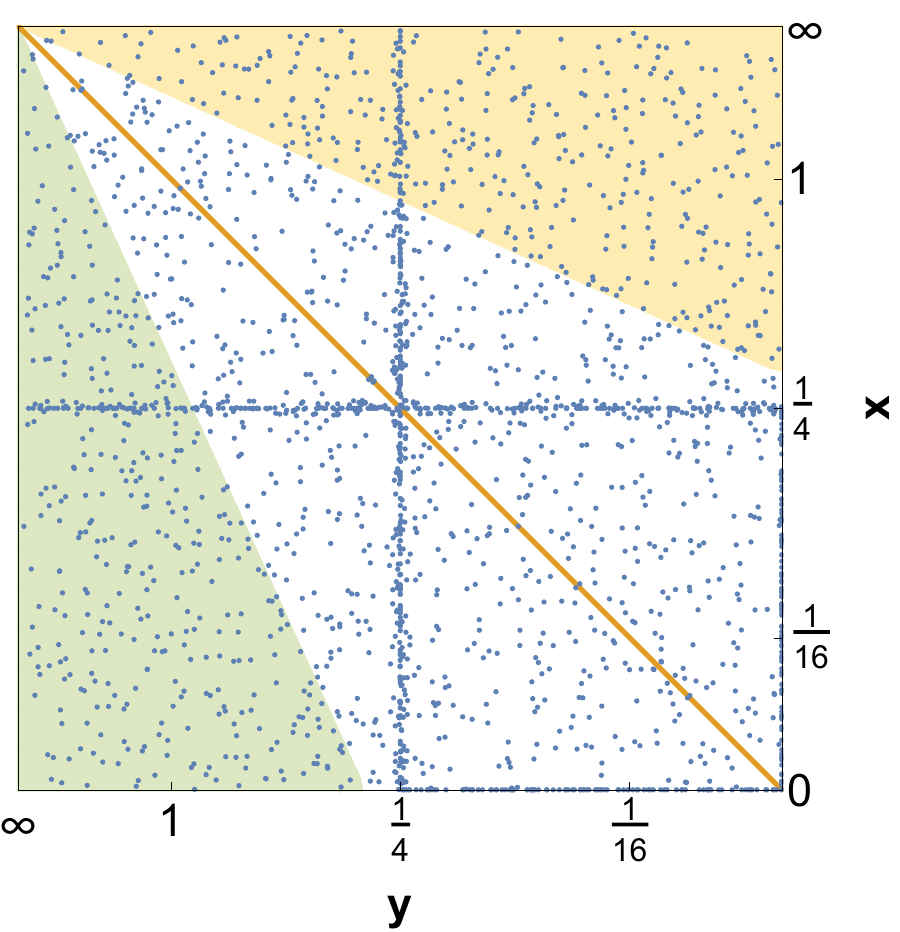}
\caption{A scatter plot displaying generated numerical samples of the three-scale coefficients. The orange line indicates the limit $x=y$. The regions highlighted in yellow and green can be accessed via expansions as explained in the text. The plot is rotated to match the perspective in Fig.~\ref{fig:results:ggHtb}.}
\label{fig:calc:grid}
\end{figure}

It is important to stress that the radii of convergence of the series expansions in Eq.~\eqref{eq:calc:exp} and Eq.~\eqref{eq:calc:coeff} are limited due to the previously imposed hierarchy of scales and the presence of physical thresholds. More precisely, expansions in the limit of large internal mass at $x\to\infty$ are only justified in the region $x>1/4$ due to the quark-pair threshold being located at $x=1/4$. Moreover, expanding the vertex-integrals encoded in the $y$-dependent expansion coefficients in Eq.~\eqref{eq:calc:coeff} implies another restriction related to physical thresholds of the corresponding two-scale remainder in Eq.~\eqref{eq:calc:LME}. Depending in the particular form factor coefficient, the exact position of the threshold varies, since not all vertex-functions enter both three-scale coefficients. For the coefficient $C^{(2)}_{tb}$, we observe the appearance of a threshold at $y=1/16$ stemming from the aforementioned two crossed vertex-integrals. Thus, the validity of the expansion in $y\to 0$ in restricted to $y<1/16$. Contrary to that, the expansion of $C^{(2)}_{bt}$ converges for $y<1/4$ due to another quark-pair threshold. The constraint on $y$ can be relaxed by inserting the exact solutions for the reduced vertex-functions. In order to fully exhaust the potential of the deep asymptotic expansions for $x\to\infty$, we employ our numerical solutions for the relevant vertex functions and sample the form factor coefficients even beyond the restricted radii of convergence in $y$ described before. In Fig.~\ref{fig:calc:grid}, the region covered by expanding $C^{(2)}_{tb}(x,y)$ at $x\to\infty$ is highlighted in yellow, whereas the corresponding region covered by expanding $C^{(2)}_{bt}(x,y)$ at $x\to\infty$ and utilizing the symmetry relation in Eq.~\eqref{eq:definition:sym} is highlighted in green. 

Owing to the previously imposed hierarchy among the mass scales, the deep asymptotic expansions in Eq.~\eqref{eq:calc:exp} of the form factor coefficients under consideration have a limited radius of convergence. Although the expansion suffices for the study of mass effects in compliance with the mass hierarchy, another technique must be adopted in order to cover the remaining regions of the two-dimensional parameter space. For this purpose, we decide to utilize the auxiliary mass flow method~\cite{Liu:2017jxz,Liu:2022mfb} as implemented in the package \texttt{AMFlow}~\cite{Liu:2022chg} and compute the master integrals at fixed numerical pairs $(x,y)$. Thereby, we not only sample the form factor coefficients in regions of the parameter space beyond the radius of convergence of the aforementioned series expansions, but also validate the quality of the expansions within their radii of convergence. More precisely, the numerical samples generated with \texttt{AMFlow} and analytic methods agree by at least ten digits in the radii of convergence of the expansions. In its default setup, \texttt{AMFlow} dresses the massive propagators with the Feynman causality parameter $i\eta$ and transports boundary conditions from the limit $\eta\to\infty$ to $\eta=0$ at fixed values of the pair $(x,y)$. Demanding a precision goal of 30 digits for the individual $\varepsilon$-expansion coefficients of the master integrals, we noticed that more than half of the computation time is spent on the construction of the underlying systems of differential equations in $\eta$, hence, on the reduction at fixed $(x,y)$ and symbolic $\eta$. We overcome this bottleneck by performing the required reductions only once with symbolic $(x,y)$ and load the constructed differential equations when using \texttt{AMFlow} to compute the master integrals point by point. Thereby the numerical calculation receives a significant speedup of more than 100\%. As far as the computation time is concerned, a single phase space point is obtained within approximately 2 hours on a laptop. 

In Fig.~\ref{fig:calc:grid}, we visualize the grid of collected numerical samples. In total, the set of master integrals has been probed in more than 2000 different points with \texttt{AMFlow} as indicated by the blue dots. For the construction of the grid, we have compactified the physical parameter space $(x,y) \in (0,\infty)^2$ to the unit square with the following transformation: 
\begin{equation} \label{eq:calc:mapping}
(x,y) \mapsto \left( \frac{1}{4}\frac{\rho}{1-\rho} , \frac{1}{4}\frac{\sigma}{1-\sigma} \right) \,.
\end{equation}
In particular, pairs in $(\rho,\sigma)\in (0,1)^2$ are uniformly distributed. In order to adequately describe possible features close to the thresholds and the limit of small mass, additional samples have been produced in the corresponding regions. 

\section{Results} \label{sect:results}

\begin{figure}[t]
\center
\includegraphics[width=0.49\textwidth]{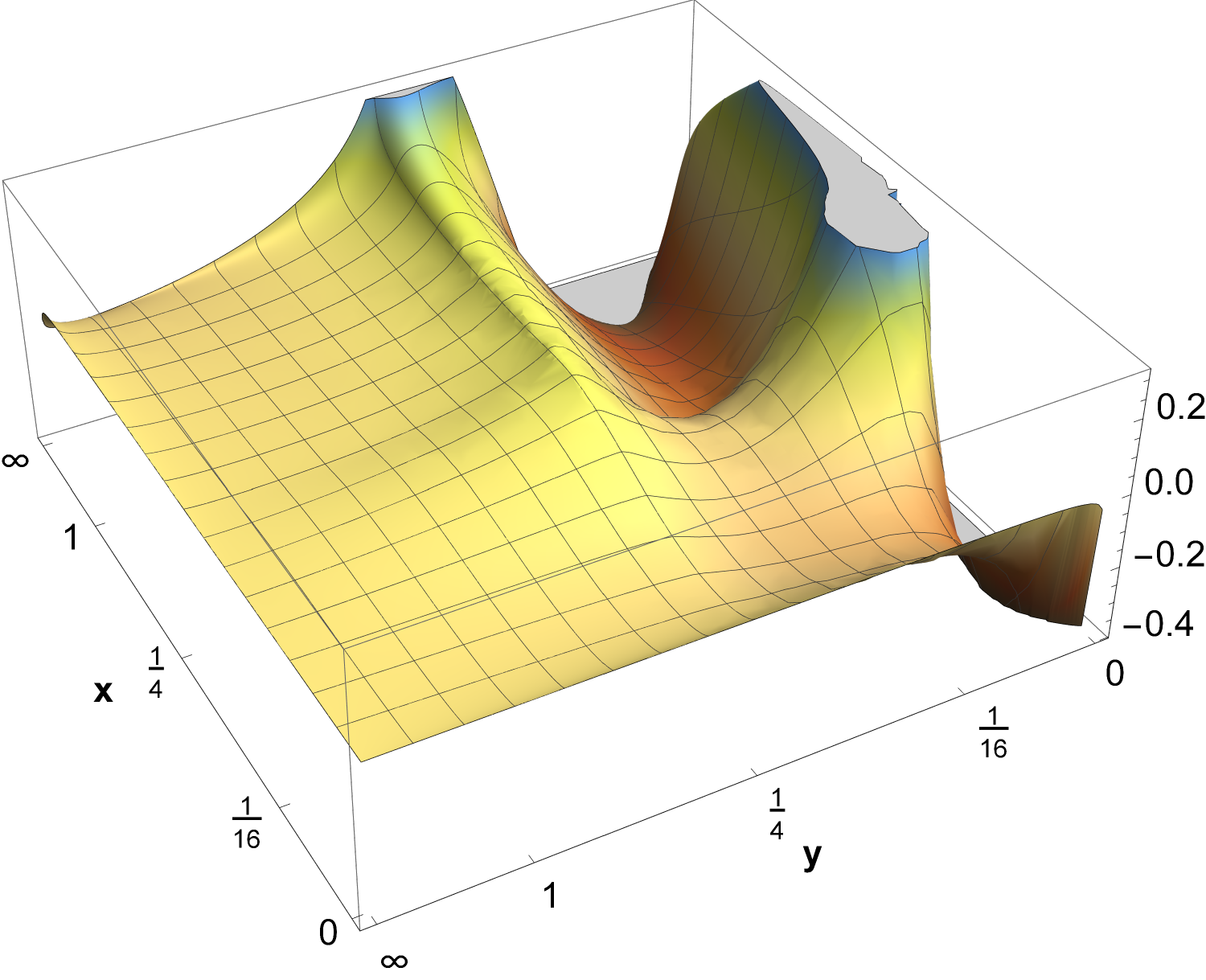}
\includegraphics[width=0.49\textwidth]{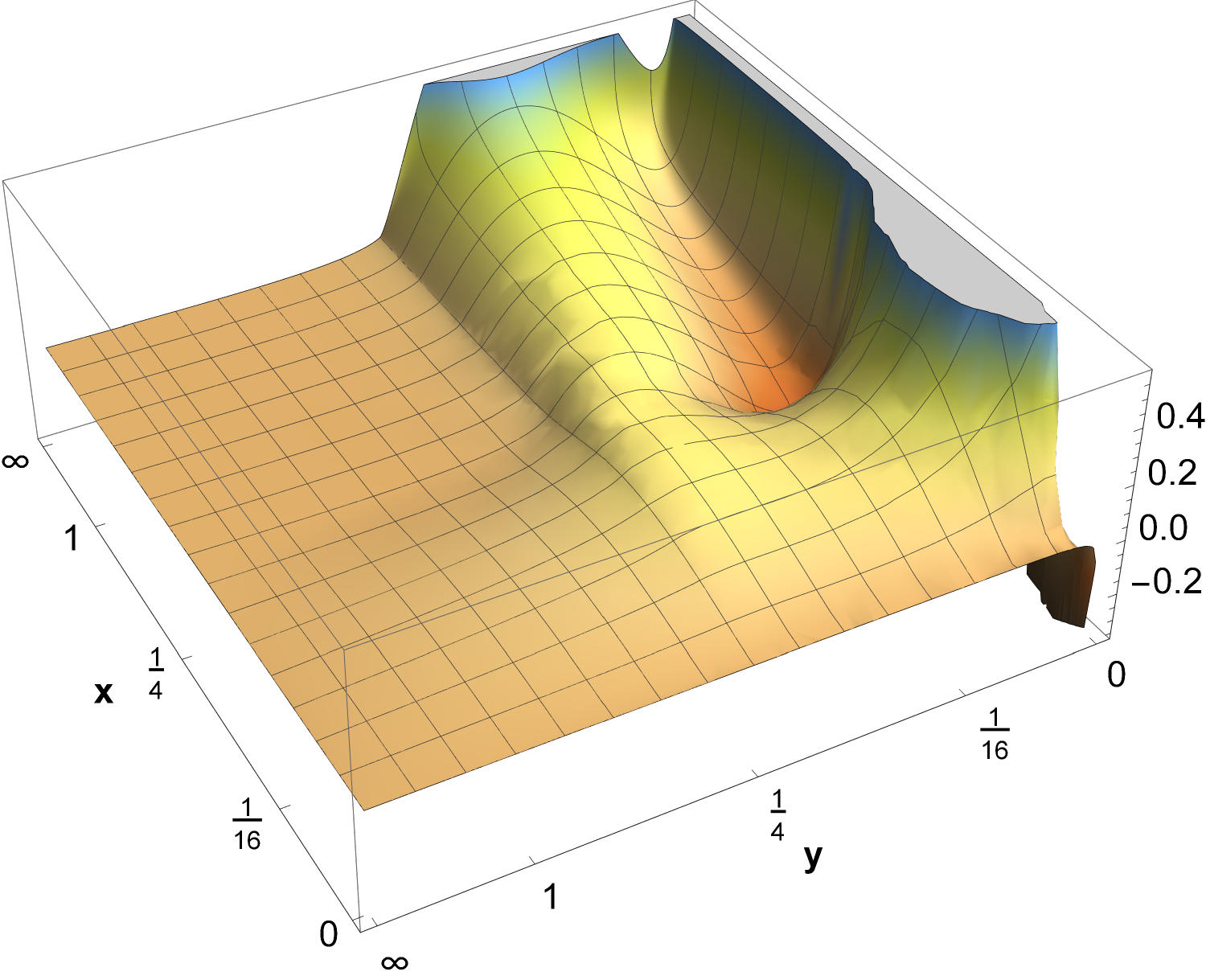}
\caption{The coefficient $\mathcal{C}^{(2)}_{tb}$ is parametrized in terms of $x$ and $y$ as defined in Eq.~\eqref{eq:definition:variables}. The left plot features the real part and the right plot shows the corresponding imaginary part. In order to visualize various features of the illustrated coefficient, we confine the vertical axis to the indicated ranges.}
\label{fig:results:ggHtb}
\end{figure}

\begin{figure}[t]
  \centering
  \begin{subfigure}[b]{0.5\textwidth}
    \centering
    \includegraphics[scale=0.45]{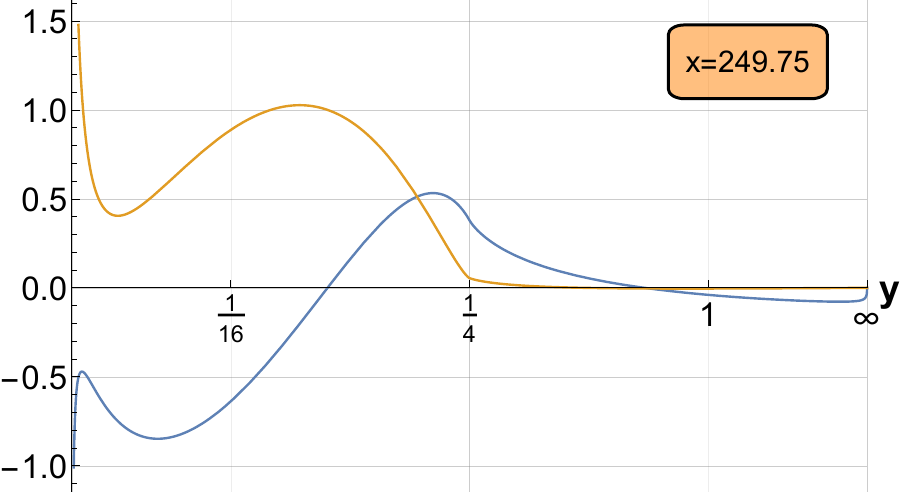}
    \caption{} \label{fig:results:slices:a}
  \end{subfigure}%
  \begin{subfigure}[b]{0.5\textwidth}
    \centering
    \includegraphics[scale=0.45]{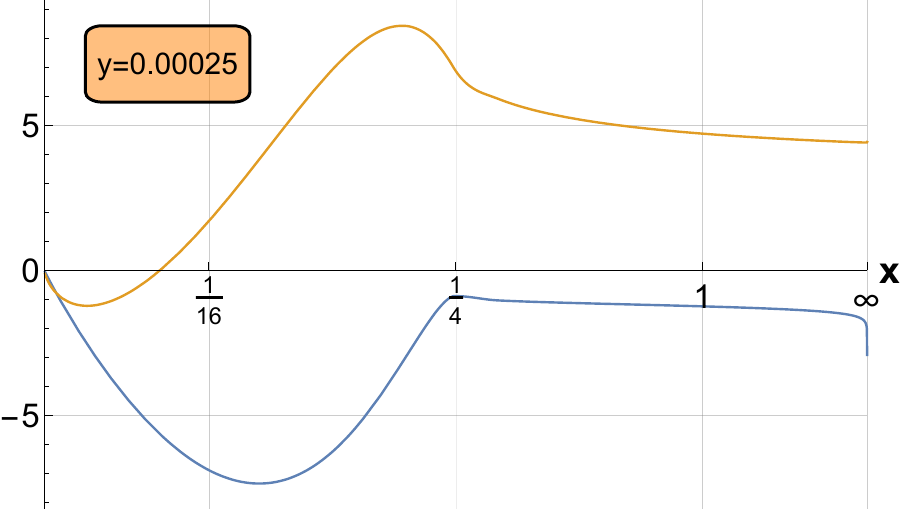}
    \caption{} \label{fig:results:slices:b}
  \end{subfigure}%
  \\
  \begin{subfigure}[b]{0.5\textwidth}
    \centering
    \includegraphics[scale=0.45]{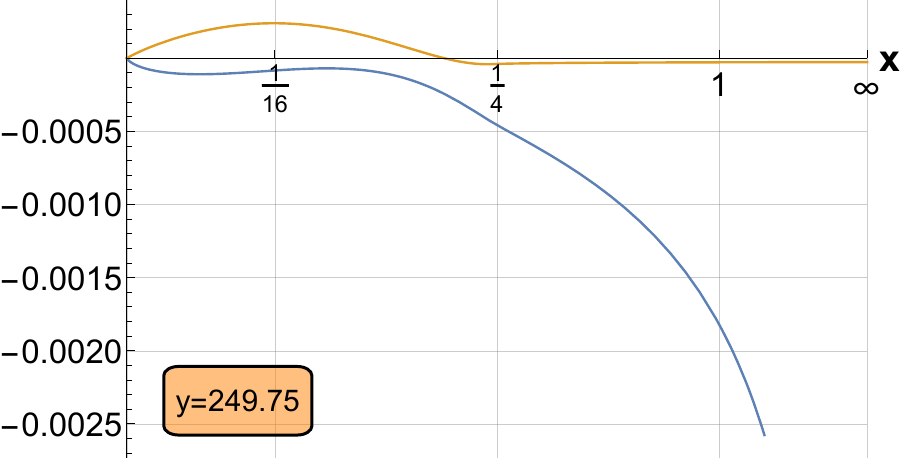}
    \caption{} \label{fig:results:slices:c}
  \end{subfigure}%
  \begin{subfigure}[b]{0.5\textwidth}
    \centering
    \includegraphics[scale=0.45]{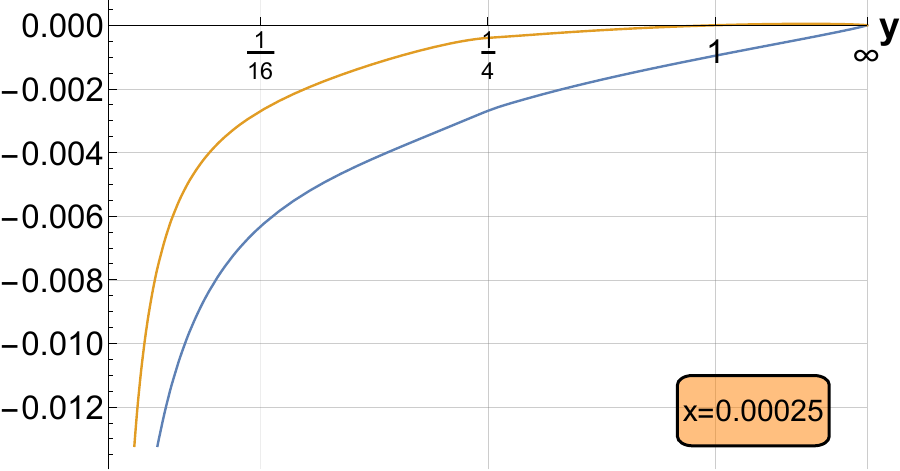}
    \caption{} \label{fig:results:slices:d}
  \end{subfigure}%
  \\
  \begin{subfigure}[b]{0.5\textwidth}
    \centering
    \includegraphics[scale=0.45]{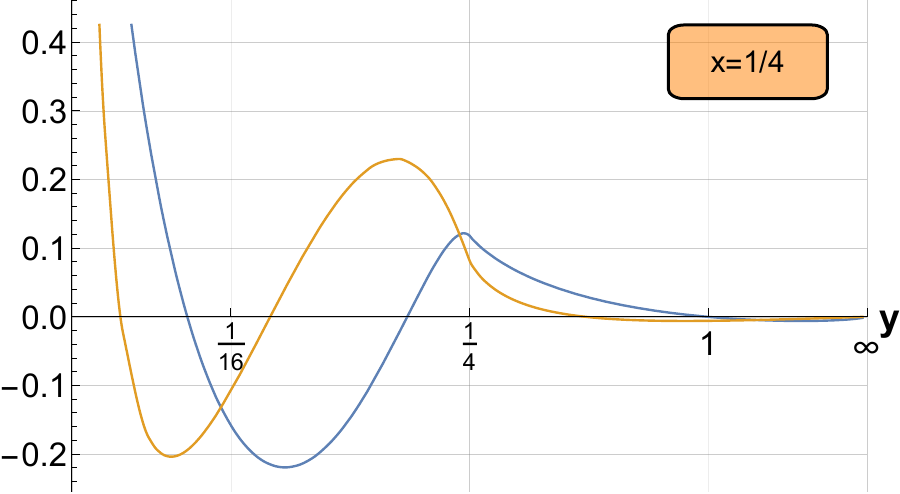}
    \caption{} \label{fig:results:slices:e}
  \end{subfigure}%
  \begin{subfigure}[b]{0.5\textwidth}
    \centering
    \includegraphics[scale=0.45]{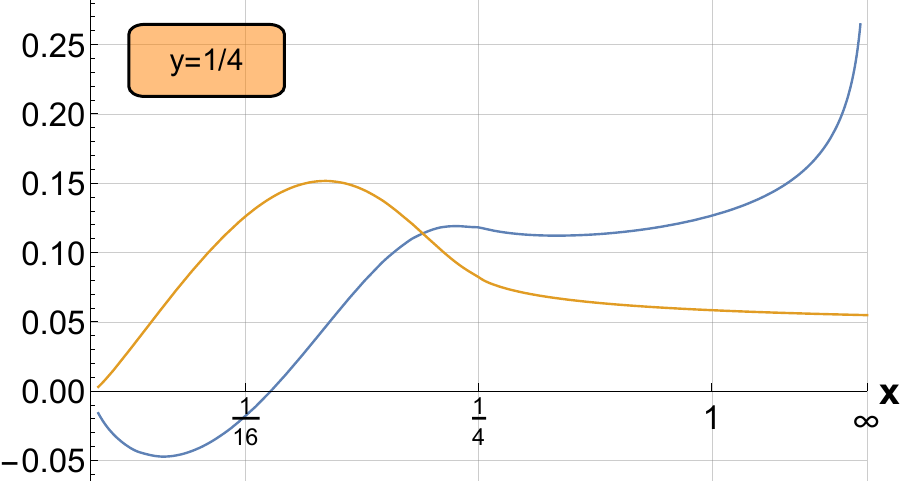}
    \caption{} \label{fig:results:slices:f}
  \end{subfigure}%
  \\
  \begin{subfigure}[b]{0.5\textwidth}
    \centering
    \includegraphics[scale=0.45]{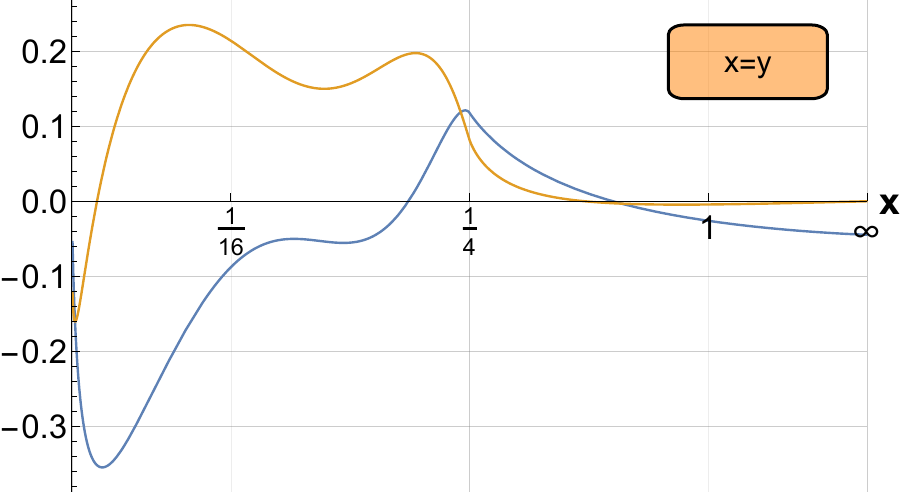}
    \caption{} \label{fig:results:slices:g}
  \end{subfigure}%
  \caption{A set of slices through the three-dimensional visualization of the coefficient $\mathcal{C}^{(2)}_{tb}$ at fixed values for $x$ or $y$ as indicated in the individual subfigures. The blue and orange curves represent the real and imaginary part, respectively.}
    \label{fig:results:slices}
\end{figure}

From the numerical samples of the master integrals obtained via \texttt{AMFlow} and the deep asymptotic expansions, we construct the bare three-scale coefficients. Their renormalization and subtraction of IR divergences yielding finite remainders is carried out in accordance with the prescriptions defined in Sect.~\ref{sect:definition}. The observation that all poles in $\varepsilon$ of the bare results cancel against those of the counterterms serves as a first consistency check of our calculation. Since the three-scale coefficients are related by the symmetry relation in Eq.~\eqref{eq:definition:sym}, we show results only for the finite remainder coefficient $\mathcal{C}^{(2)}_{tb}$. The real and imaginary part, respectively, are depicted in Fig.~\ref{fig:results:ggHtb}. To give insight into different kinematic limits, we sliced the three-dimensional illustrations by fixing one of the two variables to a constant value as presented in several subfigures of Fig.~\ref{fig:results:slices}. 

In the following, we want to discuss the finite remainder in different regions of the physical parameter space and point out some details. To facilitate a firm understanding of the limits accessible via expansions as explained in Sect.~\ref{sect:calc}, we provide the asymptotic estimate of the finite remainder $\mathcal{C}^{(2)}_{tb}$ including the leading and subleading behavior. For convenience, the exact expansion coefficients are replaced by the corresponding numerical values. In the double-limit $x\to\infty$ and $y\to 0$, the finite remainder is given by 
\begin{align} \label{eq:results:ctbXy}
\begin{autobreak}
\mathcal{C}^{(2)}_{tb}
(x,y) =
(-2.09396 
- 1.00673 i)
+ (-0.483497 
+ 0.436332 i) \, L_y
+ (0.0694444 
+ 0.1309 i) \, L_y^2
+ 0.0138889 \, L_y^3
+ \Bigl( (12.2339 
+ 10.1807 i)
+ (0.654855 
- 3.68795 i) \, L_y 
- (0.769726 
- 1.1781 i) \, L_y^2
+ (0.180556 
+ 0.0581776 i) \, L_y^3
+ 0.00462963 \, L_y^4 \Bigr) \, y
+ \Bigl[ -0.111111
+ \Bigl( 2.3116 
- 1.0472 i \, L_y
- 0.166667 \, L_y^2 \Bigr) \, y \Bigr] \, L_x
+ \biggl[ - (0.0935291 
+ 0.0296169 i)
- (0.0351998 
- 0.0254527 i) \, L_y
+ (0.00405093 
+ 0.00763582 i) \, L_y^2
+ 0.000810185 \, L_y^3
+ \Bigl( - (0.072005 
+ 0.238047 i)
- (0.132393 
+ 0.293816 i) \, L_y
- (0.0542988 
+ 0.00327249 i) \, L_y^2
+ (0.00176183 
+ 0.0033937 i) \, L_y^3
+ 0.000270062 \, L_y^4 \Bigr) \, y
+ \Bigl[ 0.00978009
+ \Bigl( 0.146737
- (0.00625 
+ 0.0397547 i) \, L_y
- 0.00632716 \, L_y^2 \Bigr) \, y \Bigr] \, L_x
+ 0.003125 \, y \, L_x^2 \biggr] \, \frac{1}{x}
+ \mathcal{O}\left(\frac{1}{x^{3/2}},\, y^2\right) \,,
\end{autobreak}
\end{align}
with $L_x=\log(x)$ and $L_y=\log(y)$. We note that the leading asymptotic behavior starts contributing at constant order in $x$ and $y$. Furthermore, it is worth to mention that beyond the subleading power, square roots in $x$ and $y$ start to enter the expansion. Their appearance can be traced back to the renormalization of the quark masses, since the expansion of the bare form factors as well as the other renormalization constants do not involve roots. Contrary to that limit, the asymptotic estimate for $y\to\infty$ and $x\to 0$ takes the form 
\begin{align} \label{eq:results:ctbYx}
\begin{autobreak}
\mathcal{C}^{(2)}_{tb}
(x,y) =
 \biggl[ \Bigl( (0.559207 
 + 0.288409 i)
+ (0.225378 
- 0.239183 i) \, L_x
- (0.0396296 
+ 0.0349066 i) \, L_x^2 \Bigr) \, x
+ \Bigl( - (2.75692 
+ 4.52439 i)
- (2.63376 
- 1.44643 i) \, L_x
+ (0.249304 
+ 0.523053 i) \, L_x^2
+ 0.0269097 \, L_x^3 \Bigr) \, x^{2}
+ \Bigl[ \Bigl( (0.0857502 
- 0.00981748 i)
+ (0.003125 
- 0.0698132 i) \, L_x
- 0.0111111 \, L_x^2 \Bigr) \, x
+ \Bigl( - (1.03048 
+ 0.119991 i)
- (0.0381944 
- 0.53887 i) \, L_x
+ 0.0857639 \, L_x^2 \Bigr) \, x^{2} \Bigr] \, L_y
- 0.003125 \, x\, L_y^2 \biggr] \, \frac{1}{y}
+ \biggl[ \Bigl( (0.00408711 
+ 0.0638994 i)
- (0.0136805 
- 0.00864205 i) \, L_x
+ (0.00149943 
+ 0.00187 i) \, L_x^2 \Bigr) \, x
+ \Bigl( (0.243953 
+ 0.097041 i)
+ (0.209498 
- 0.0574337 i) \, L_x
- (0.0211944 
+ 0.0274266 i) \, L_x^2 \Bigr) \, x^{2}
+ \Bigl[ \Bigl( (0.0222708 
+ 0.0285174 i)
- (0.000248016 
- 0.00373999 i) \, L_x
+ 0.000595238 \, L_x^2 \Bigr) \, x
+ \Bigl( - (0.00628293 
+ 0.167365 i)
+ (0.0241071 
- 0.0548532 i) \, L_x
- 0.00873016 \, L_x^2 \Bigr) \, x^{2} \Bigr] \, L_y
+ \Bigl[ 0.0046627 \, x
- 0.0386905 \, x^{2} \Bigr] \, L_y^2 \biggr] \, \frac{1}{y^2}
+ \mathcal{O}\left(\frac{1}{y^3},\, x^3\right) \,.
\end{autobreak}
\end{align}
Here, the finite remainder is power-suppressed in both variables. Hence, the finite remainder vanishes in this limit as expected for a massless quark flavor coupling to the Higgs-boson. As opposed to the previous limit, contributions proportional to square roots in $x$ or $y$ due to renormalization of the quark masses cancel out completely. 

To the best of our knowledge, the contribution to the Higgs-gluon form factor under consideration has not been addressed in the literature so far. Thus, performing independent cross-checks of our results, for example, in special kinematic limits is possible only in a single case. The only result to compare with is given by the limit of equal quark masses being covered in Ref.~\cite{Czakon:2020vql}. Starting from general values in the two-dimensional parameter space, the corresponding limit $x=y$ can be smoothly reached, where we find perfect agreement with previous findings. However, it must be mentioned that this evidence does not provide the strongest check for the correctness of our results, since only the limit but not the neighborhood is probed. In Fig.~\ref{fig:results:slices:g}, the corresponding limit of the finite remainder is depicted. It is not difficult to convince ourselves that the finite remainder vanishes in $x=y=0$ due to the well-known subleading power behavior. Furthermore, we note that the imaginary part disappears for large $x$ and $y$, while the real part approaches the constant $-77/1728$. Between these boundaries, the finite remainder is a continuously differentiable function and varies smoothly at the physical thresholds. 

Beyond the limit of equal quark masses, the finite remainder is most conveniently analyzed in the regions of large scale hierarchy, where the expansions in Eq.~\eqref{eq:results:ctbXy} for $x\gg y$ and Eq.~\eqref{eq:results:ctbYx} for $y\gg x$, respectively, are valid. At first sight, it is striking that the finite remainder as presented in Fig.~\ref{fig:results:ggHtb} develops a divergent asymptotic behavior for small values of $y$ below the threshold located at $y=1/4$ independent of $x>0$, whereas on the other side of the threshold for $y>1/4$, the finite remainder is relatively flat. According to the asymptotic expansion in Eq.~\eqref{eq:results:ctbXy}, the divergent behavior of the finite remainder is of logarithmic origin, while in the opposite limit, the flat landscape can be attributed to the power-suppressed character of the form factor as formulated in Eq.~\eqref{eq:results:ctbYx}. As far as the power-suppression is concerned, it is clear that the finite remainder approaches zero for $x\to 0$, being in compliance with the expectation that the form factor vanishes when the Higgs-boson couples to a massless quark. Similarly, the finite remainder goes to zero for $y\to\infty$ and $x<\infty$. We note that the region for $y\to\infty$ becomes flat only after renormalization. In fact, the bare three-scale coefficient exhibits a logarithmic divergence at the finite order in $y$ being completely canceled via renormalization. 

Special care must be taken when considering the transition of the real part for $y\to\infty$ from $x\neq y$ to the limit where both masses become infinitely large at the same time. The finite remainder vanishes in the addressed regions, which is underlined with the slices in Figs.~\ref{fig:results:slices:c}~,~\ref{fig:results:slices:d}~and~\ref{fig:results:slices:e}. The Figs.~\ref{fig:results:slices:a}~and~\ref{fig:results:slices:c} consistently showcase the vanishing of the imaginary part in agreement with the known limit of equal mass, where $x$ and $y$ approach infinity at the same rate. The corresponding real part converges much slower to the non-zero constant as stated above and seems to experience a sharp turn when one parameter is kept finite while approaching infinity with the other one. The intricate asymptotics of the real part in the neighborhood of the boundary point $1/x=1/y=0$ are stressed in Figs.~\ref{fig:results:slices:a}~and~\ref{fig:results:slices:c}. More precisely, in Fig.~\ref{fig:results:slices:a}, we observe that the real part converges to zero for $y\to\infty$ and large but finite $x$, whereas in Fig.~\ref{fig:results:slices:c}, the real part diverges at large fixed $y$ for $x\to\infty$. With the methods at our disposal, it is not possible to further analyze the precise transition from $x=y$ at infinitely large values to large but finite $x$ or $y$ independently, due to a lack of scale hierarchy in that region. Likewise, the same difficulty applies to the neighborhood of the limit where $x$ and $y$ tend to zero at potentially different rates. The Figs.~\ref{fig:results:slices:b}~and~\ref{fig:results:slices:d} highlight the problem at hand. While in Fig.~\ref{fig:results:slices:b}, real and imaginary part uniformly converge to zero for $x\to 0$ as demanded by the limit of equal mass, Fig.~\ref{fig:results:slices:d} displays singular asymptotics for finite $x$ as $y\to 0$. 

In contrast to the flat region of the finite remainder, the logarithmically enhanced region seems to induce sizable corrections as far as the three-scale coefficient is considered as a virtual correction in hadronic Higgs production. For small values of $y$, the corresponding quark flavor running in the loop which is not directly coupled to the Higgs-boson becomes effectively massless. It is important to recall that if we had assigned a massless quark flavor to the second fermion loop from the outset, its effect would have been regularized in terms of poles in the dimensional regulator. Since the singularity is now screened by the mass of the quark, a logarithmic divergence is developed. We note that the logarithms in the mass of the second quark flavor cannot be absorbed by a redefinition of the strong coupling constant. Hence, in the phenomenological application a consistent treatment within the 4-flavor scheme is inevitable, since the logarithms in the mass are supposed to cancel against logarithms stemming from gluon splittings into massive final-state quarks in real corrections by virtue of the KLN theorem \cite{Kinoshita:1962ur,Lee:1964is}. Clearly, the impact of the logarithmic divergence is maximized for a large separation of scales and for the heavier quark directly coupled to the Higgs-boson to prevent a suppression due to the Yukawa coupling. Logarithmic grows of the finite remainder in the aforementioned regions are visible in the slices in Figs.~\ref{fig:results:slices:a}~,~\ref{fig:results:slices:b}~,~\ref{fig:results:slices:e}~and~\ref{fig:results:slices:f}. In this context, it is worth to note that for fixed $y$, only the real part is subject to the logarithmic growth in the limit of large $x$. Contrary to the real part, the imaginary part assumes a constant value in the limit as clearly visible in Figs.~\ref{fig:results:slices:b}~and~\ref{fig:results:slices:f}. This property can also be deduced from Eq.~\eqref{eq:results:ctbXy}, in particular, the leading asymptotic behavior. 

\section{Conclusions and outlook} \label{sect:conclusion}

We completed the calculation of the Higgs-gluon form factor at three loops in QCD with multiple massive quark flavors. In particular, three-loop corrections to the form factor give rise to a class of Feynman diagrams involving two closed loops of possibly different massive quark flavors. Since the exact three-loop computation of Ref.~\cite{Czakon:2020vql} relies on QCD with a single massive quark flavor, the aforementioned subset of Feynman diagrams was discarded. With the findings presented in this paper, Ref.~\cite{Czakon:2020vql} is supplemented with the missing contribution such that the Higgs-gluon form factor is known for arbitrary configurations of mass parameters. 

The contribution originating from Feynman diagrams with two different massive quark flavors is parametrized in terms of two dimensionless variables. In order to describe the contribution to the form factor, we employed two complementary methods both of which rest on the method of differential equations. On the one hand, we constructed deep asymptotic expansions of the underlying master integrals in limits relevant for the phenomenological application. On the other hand, the regions of the parameter space beyond the radii of convergence of the asymptotic expansions were sampled by numerically solving the relevant master integrals with \texttt{AMFlow}. Thereby, the physical region of the parameter space is covered by a combination of analytic expansions and numerical samples. We provide an ancillary file with all results obtained in the course of this work as a supplement to this paper. Additionally, we agree to generate more numerical samples of the form factor upon request. Since our contribution to the form factor is presented as a finite remainder, we refer to the analytic lower-order results known to higher orders in the dimensional regulator~\cite{Anastasiou:2020qzk}, which may be used to reconstruct the IR poles or to change the renormalization scheme for the mass parameters. 

In combination with the results of Ref.~\cite{Czakon:2020vql}, our findings constitute a main building block for the evaluation of the top-bottom interference effects in Higgs production via gluon fusion at NNLO in QCD. In light of the phenomenological application, we want to stress that for a consistent treatment in 4-flavor scheme, the new contribution to the form factor is indispensable as it cancels against large logarithmic corrections due to gluon splittings into massive bottom quarks in real radiation. 

Finally, we point out the possibility to apply the methods employed in the course of our work to the calculation of similar form factors, for example, the Higgs-photon form factor or massless quark form factors at the tree-loop level in QCD with two different massive quark flavors. Since the Higgs-photon form factor as well as the massless quark form factors share large subsets of relevant integrals with those encountered in the Higgs-gluon form factor, their evaluation is feasible upon minor modifications. We dedicate the computation to future work. 

\section*{Acknowledgements}

We are indebted to Micha\l{} Czakon for generously providing access to the private software package \texttt{DiaGen/IdSolver} that enhanced the workflow of our calculation. All Feynman diagrams have been drawn using the tool \texttt{FeynGame}~\cite{Harlander:2020cyh}. 

\newpage

\appendix

\section{Supplemental material} \label{app:supplemental}

A collection of our results for the coefficient $C^{(2)}_{tb}$ can be found in the ancillary file~\cite{mathfile} written in \textsc{Wolfram} language. The following variables are defined for deep asymptotic expansions with exact coefficients: 
\begin{description}
\item \verb|CtbXy| - Expansion of $C^{(2)}_{tb}(x,y)$ in the limit $x\to\infty$ and $y\to 0$ (cf. Eq.~\eqref{eq:results:ctbXy});
\item \verb|CtbYx| - Expansion of $C^{(2)}_{tb}(x,y)$ in the limit $y\to\infty$ and $x\to 0$ (cf. Eq.~\eqref{eq:results:ctbYx}).
\end{description}
As pointed out in Sect.~\ref{sect:calc}, the corresponding radii of convergence are limited due to the presence of thresholds. More precisely, $C^{(2)}_{tb}(x,y)$ expanded in the double-limit $x\to\infty$ and $y\to 0$ is convergent under the constraints $x>1/4$, $y<1/16$ and $y\ll x$, whereas the expansion in the opposite limit $y\to\infty$ and $x\to 0$ can be applied for $y>1/4$, $x<1/4$ and $x\ll y$. Additionally, grids with numerical samples for $C^{(2)}_{tb}$ are included: 
\begin{description}
\item \verb|pointsAMF| - 2294 samples distributed according to the blue dots in Fig.~\ref{fig:calc:grid};
\item \verb|pointsLME| - 50000 samples obtained via expansions in the colored regions in Fig.~\ref{fig:calc:grid};
\item \verb|pointsEqM| - 999 samples at $x=y$ extracted from \cite{Czakon:2020vql}.
\end{description}
We recall that in the grids of numerical samples, the physical domain $(x,y)\in(0,\infty)^2$ of the parameter space has been compactified to the unit square in accordance with Eq.~\eqref{eq:calc:mapping}. Hence, the first two fields in the grids correspond to values in $\rho$ and $\sigma$. The same conformal mapping has been applied to generate the illustrations in Sect.~\ref{sect:results}. 

\newpage

\bibliographystyle{JHEP}
\bibliography{ggH}

\providecommand{\href}[2]{#2}\begingroup\raggedright\begin{thebibliography}{10}

\bibitem{ATLAS:2012yve}
{\scshape ATLAS} collaboration, \emph{{Observation of a new particle in the
  search for the Standard Model Higgs boson with the ATLAS detector at the
  LHC}}, \href{https://doi.org/10.1016/j.physletb.2012.08.020}{\emph{Phys.
  Lett. B} {\bfseries 716} (2012) 1}
  [\href{https://arxiv.org/abs/1207.7214}{{\ttfamily 1207.7214}}].

\bibitem{CMS:2012qbp}
{\scshape CMS} collaboration, \emph{{Observation of a New Boson at a Mass of
  125 GeV with the CMS Experiment at the LHC}},
  \href{https://doi.org/10.1016/j.physletb.2012.08.021}{\emph{Phys. Lett. B}
  {\bfseries 716} (2012) 30} [\href{https://arxiv.org/abs/1207.7235}{{\ttfamily
  1207.7235}}].

\bibitem{ATLAS:2022vkf}
{\scshape ATLAS} collaboration, \emph{{A detailed map of Higgs boson
  interactions by the ATLAS experiment ten years after the discovery}},
  \href{https://doi.org/10.1038/s41586-022-04893-w}{\emph{Nature} {\bfseries
  607} (2022) 52} [\href{https://arxiv.org/abs/2207.00092}{{\ttfamily
  2207.00092}}].

\bibitem{CMS:2022dwd}
{\scshape CMS} collaboration, \emph{{A portrait of the Higgs boson by the CMS
  experiment ten years after the discovery.}},
  \href{https://doi.org/10.1038/s41586-022-04892-x}{\emph{Nature} {\bfseries
  607} (2022) 60} [\href{https://arxiv.org/abs/2207.00043}{{\ttfamily
  2207.00043}}].

\bibitem{LHCHiggsCrossSectionWorkingGroup:2016ypw}
{\scshape LHC Higgs Cross Section Working Group} collaboration, \emph{{Handbook
  of LHC Higgs Cross Sections: 4. Deciphering the Nature of the Higgs Sector}},
   \href{https://arxiv.org/abs/1610.07922}{{\ttfamily 1610.07922}}.

\bibitem{Spira:1995rr}
M.~Spira, A.~Djouadi, D.~Graudenz and P.M.~Zerwas, \emph{{Higgs boson
  production at the LHC}},
  \href{https://doi.org/10.1016/0550-3213(95)00379-7}{\emph{Nucl. Phys. B}
  {\bfseries 453} (1995) 17}
  [\href{https://arxiv.org/abs/hep-ph/9504378}{{\ttfamily hep-ph/9504378}}].

\bibitem{Harlander:2005rq}
R.~Harlander and P.~Kant, \emph{{Higgs production and decay: Analytic results
  at next-to-leading order QCD}},
  \href{https://doi.org/10.1088/1126-6708/2005/12/015}{\emph{JHEP} {\bfseries
  12} (2005) 015} [\href{https://arxiv.org/abs/hep-ph/0509189}{{\ttfamily
  hep-ph/0509189}}].

\bibitem{Anastasiou:2006hc}
C.~Anastasiou, S.~Beerli, S.~Bucherer, A.~Daleo and Z.~Kunszt, \emph{{Two-loop
  amplitudes and master integrals for the production of a Higgs boson via a
  massive quark and a scalar-quark loop}},
  \href{https://doi.org/10.1088/1126-6708/2007/01/082}{\emph{JHEP} {\bfseries
  01} (2007) 082} [\href{https://arxiv.org/abs/hep-ph/0611236}{{\ttfamily
  hep-ph/0611236}}].

\bibitem{Aglietti:2006tp}
U.~Aglietti, R.~Bonciani, G.~Degrassi and A.~Vicini, \emph{{Analytic Results
  for Virtual QCD Corrections to Higgs Production and Decay}},
  \href{https://doi.org/10.1088/1126-6708/2007/01/021}{\emph{JHEP} {\bfseries
  01} (2007) 021} [\href{https://arxiv.org/abs/hep-ph/0611266}{{\ttfamily
  hep-ph/0611266}}].

\bibitem{Harlander:2009bw}
R.V.~Harlander and K.J.~Ozeren, \emph{{Top mass effects in Higgs production at
  next-to-next-to-leading order QCD: Virtual corrections}},
  \href{https://doi.org/10.1016/j.physletb.2009.08.012}{\emph{Phys. Lett. B}
  {\bfseries 679} (2009) 467}
  [\href{https://arxiv.org/abs/0907.2997}{{\ttfamily 0907.2997}}].

\bibitem{Pak:2009bx}
A.~Pak, M.~Rogal and M.~Steinhauser, \emph{{Virtual three-loop corrections to
  Higgs boson production in gluon fusion for finite top quark mass}},
  \href{https://doi.org/10.1016/j.physletb.2009.08.016}{\emph{Phys. Lett. B}
  {\bfseries 679} (2009) 473}
  [\href{https://arxiv.org/abs/0907.2998}{{\ttfamily 0907.2998}}].

\bibitem{Davies:2019wmk}
J.~Davies, F.~Herren and M.~Steinhauser, \emph{{Top Quark Mass Effects in Higgs
  Boson Production at Four-Loop Order: Virtual Corrections}},
  \href{https://doi.org/10.1103/PhysRevLett.124.112002}{\emph{Phys. Rev. Lett.}
  {\bfseries 124} (2020) 112002}
  [\href{https://arxiv.org/abs/1911.10214}{{\ttfamily 1911.10214}}].

\bibitem{Grober:2017uho}
R.~Gr\"ober, A.~Maier and T.~Rauh, \emph{{Reconstruction of top-quark mass
  effects in Higgs pair production and other gluon-fusion processes}},
  \href{https://doi.org/10.1007/JHEP03(2018)020}{\emph{JHEP} {\bfseries 03}
  (2018) 020} [\href{https://arxiv.org/abs/1709.07799}{{\ttfamily
  1709.07799}}].

\bibitem{Davies:2019nhm}
J.~Davies, R.~Gr\"ober, A.~Maier, T.~Rauh and M.~Steinhauser, \emph{{Top quark
  mass dependence of the Higgs boson-gluon form factor at three loops}},
  \href{https://doi.org/10.1103/PhysRevD.100.034017}{\emph{Phys. Rev. D}
  {\bfseries 100} (2019) 034017}
  [\href{https://arxiv.org/abs/1906.00982}{{\ttfamily 1906.00982}}].

\bibitem{Czakon:2020vql}
M.~Czakon and M.~Niggetiedt, \emph{{Exact quark-mass dependence of the
  Higgs-gluon form factor at three loops in QCD}},
  \href{https://doi.org/10.1007/JHEP05(2020)149}{\emph{JHEP} {\bfseries 05}
  (2020) 149} [\href{https://arxiv.org/abs/2001.03008}{{\ttfamily
  2001.03008}}].

\bibitem{Harlander:2019ioe}
R.V.~Harlander, M.~Prausa and J.~Usovitsch, \emph{{The light-fermion
  contribution to the exact Higgs-gluon form factor in QCD}},
  \href{https://doi.org/10.1007/JHEP10(2019)148}{\emph{JHEP} {\bfseries 10}
  (2019) 148} [\href{https://arxiv.org/abs/1907.06957}{{\ttfamily
  1907.06957}}].

\bibitem{Prausa:2020psw}
M.~Prausa and J.~Usovitsch, \emph{{The analytic leading color contribution to
  the Higgs-gluon form factor in QCD at NNLO}},
  \href{https://doi.org/10.1007/JHEP03(2021)127}{\emph{JHEP} {\bfseries 03}
  (2021) 127} [\href{https://arxiv.org/abs/2008.11641}{{\ttfamily
  2008.11641}}].

\bibitem{Bernreuther:1981sg}
W.~Bernreuther and W.~Wetzel, \emph{{Decoupling of Heavy Quarks in the Minimal
  Subtraction Scheme}},
  \href{https://doi.org/10.1016/0550-3213(82)90288-7}{\emph{Nucl. Phys. B}
  {\bfseries 197} (1982) 228}.

\bibitem{Larin:1994va}
S.A.~Larin, T.~van Ritbergen and J.A.M.~Vermaseren, \emph{{The Large quark mass
  expansion of Gamma (Z0 ---\ensuremath{>} hadrons) and Gamma (tau-
  ---\ensuremath{>} tau-neutrino + hadrons) in the order alpha-s**3}},
  \href{https://doi.org/10.1016/0550-3213(94)00574-X}{\emph{Nucl. Phys. B}
  {\bfseries 438} (1995) 278}
  [\href{https://arxiv.org/abs/hep-ph/9411260}{{\ttfamily hep-ph/9411260}}].

\bibitem{Czakon:2007wk}
M.~Czakon, A.~Mitov and S.~Moch, \emph{{Heavy-quark production in gluon fusion
  at two loops in QCD}},
  \href{https://doi.org/10.1016/j.nuclphysb.2008.02.001}{\emph{Nucl. Phys. B}
  {\bfseries 798} (2008) 210}
  [\href{https://arxiv.org/abs/0707.4139}{{\ttfamily 0707.4139}}].

\bibitem{Gray:1990yh}
N.~Gray, D.J.~Broadhurst, W.~Grafe and K.~Schilcher, \emph{{Three Loop Relation
  of Quark (Modified) Ms and Pole Masses}},
  \href{https://doi.org/10.1007/BF01614703}{\emph{Z. Phys. C} {\bfseries 48}
  (1990) 673}.

\bibitem{Davydychev:1998si}
A.I.~Davydychev and A.G.~Grozin, \emph{{Effect of m(c) on b quark
  chromomagnetic interaction and on-shell two loop integrals with two masses}},
  \href{https://doi.org/10.1103/PhysRevD.59.054023}{\emph{Phys. Rev. D}
  {\bfseries 59} (1999) 054023}
  [\href{https://arxiv.org/abs/hep-ph/9809589}{{\ttfamily hep-ph/9809589}}].

\bibitem{Anastasiou:2020qzk}
C.~Anastasiou, N.~Deutschmann and A.~Schweitzer, \emph{{Quark mass effects in
  two-loop Higgs amplitudes}},
  \href{https://doi.org/10.1007/JHEP07(2020)113}{\emph{JHEP} {\bfseries 07}
  (2020) 113} [\href{https://arxiv.org/abs/2001.06295}{{\ttfamily
  2001.06295}}].

\bibitem{Catani:1998bh}
S.~Catani, \emph{{The Singular behavior of QCD amplitudes at two loop order}},
  \href{https://doi.org/10.1016/S0370-2693(98)00332-3}{\emph{Phys. Lett. B}
  {\bfseries 427} (1998) 161}
  [\href{https://arxiv.org/abs/hep-ph/9802439}{{\ttfamily hep-ph/9802439}}].

\bibitem{deFlorian:2012za}
D.~de~Florian and J.~Mazzitelli, \emph{{A next-to-next-to-leading order
  calculation of soft-virtual cross sections}},
  \href{https://doi.org/10.1007/JHEP12(2012)08}{\emph{JHEP} {\bfseries 12}
  (2012) 088} [\href{https://arxiv.org/abs/1209.0673}{{\ttfamily 1209.0673}}].

\bibitem{Becher:2009cu}
T.~Becher and M.~Neubert, \emph{{Infrared singularities of scattering
  amplitudes in perturbative QCD}},
  \href{https://doi.org/10.1103/PhysRevLett.102.162001}{\emph{Phys. Rev. Lett.}
  {\bfseries 102} (2009) 162001}
  [\href{https://arxiv.org/abs/0901.0722}{{\ttfamily 0901.0722}}].

\bibitem{Becher:2009qa}
T.~Becher and M.~Neubert, \emph{{On the Structure of Infrared Singularities of
  Gauge-Theory Amplitudes}},
  \href{https://doi.org/10.1088/1126-6708/2009/06/081}{\emph{JHEP} {\bfseries
  06} (2009) 081} [\href{https://arxiv.org/abs/0903.1126}{{\ttfamily
  0903.1126}}].

\bibitem{DiaGen}
M.~Czakon, ``{DiaGen/IdSolver}.''.

\bibitem{Vermaseren:2000nd}
J.A.M.~Vermaseren, \emph{{New features of FORM}},
  \href{https://arxiv.org/abs/math-ph/0010025}{{\ttfamily math-ph/0010025}}.

\bibitem{Kuipers:2012rf}
J.~Kuipers, T.~Ueda, J.A.M.~Vermaseren and J.~Vollinga, \emph{{FORM version
  4.0}}, \href{https://doi.org/10.1016/j.cpc.2012.12.028}{\emph{Comput. Phys.
  Commun.} {\bfseries 184} (2013) 1453}
  [\href{https://arxiv.org/abs/1203.6543}{{\ttfamily 1203.6543}}].

\bibitem{Ruijl:2017dtg}
B.~Ruijl, T.~Ueda and J.~Vermaseren, \emph{{FORM version 4.2}},
  \href{https://arxiv.org/abs/1707.06453}{{\ttfamily 1707.06453}}.

\bibitem{vanRitbergen:1998pn}
T.~van Ritbergen, A.N.~Schellekens and J.A.M.~Vermaseren, \emph{{Group theory
  factors for Feynman diagrams}},
  \href{https://doi.org/10.1142/S0217751X99000038}{\emph{Int. J. Mod. Phys. A}
  {\bfseries 14} (1999) 41}
  [\href{https://arxiv.org/abs/hep-ph/9802376}{{\ttfamily hep-ph/9802376}}].

\bibitem{Tkachov:1981wb}
F.V.~Tkachov, \emph{{A Theorem on Analytical Calculability of Four Loop
  Renormalization Group Functions}},
  \href{https://doi.org/10.1016/0370-2693(81)90288-4}{\emph{Phys. Lett. B}
  {\bfseries 100} (1981) 65}.

\bibitem{Chetyrkin:1981qh}
K.G.~Chetyrkin and F.V.~Tkachov, \emph{{Integration by Parts: The Algorithm to
  Calculate beta Functions in 4 Loops}},
  \href{https://doi.org/10.1016/0550-3213(81)90199-1}{\emph{Nucl. Phys. B}
  {\bfseries 192} (1981) 159}.

\bibitem{Laporta:2000dsw}
S.~Laporta, \emph{{High precision calculation of multiloop Feynman integrals by
  difference equations}},
  \href{https://doi.org/10.1142/S0217751X00002159}{\emph{Int. J. Mod. Phys. A}
  {\bfseries 15} (2000) 5087}
  [\href{https://arxiv.org/abs/hep-ph/0102033}{{\ttfamily hep-ph/0102033}}].

\bibitem{Maierhofer:2017gsa}
P.~Maierh\"ofer, J.~Usovitsch and P.~Uwer, \emph{{Kira\textemdash{}A Feynman
  integral reduction program}},
  \href{https://doi.org/10.1016/j.cpc.2018.04.012}{\emph{Comput. Phys. Commun.}
  {\bfseries 230} (2018) 99}
  [\href{https://arxiv.org/abs/1705.05610}{{\ttfamily 1705.05610}}].

\bibitem{Maierhofer:2018gpa}
P.~Maierh\"ofer and J.~Usovitsch, \emph{{Kira 1.2 Release Notes}},
  \href{https://arxiv.org/abs/1812.01491}{{\ttfamily 1812.01491}}.

\bibitem{Klappert:2020nbg}
J.~Klappert, F.~Lange, P.~Maierh\"ofer and J.~Usovitsch, \emph{{Integral
  reduction with Kira 2.0 and finite field methods}},
  \href{https://doi.org/10.1016/j.cpc.2021.108024}{\emph{Comput. Phys. Commun.}
  {\bfseries 266} (2021) 108024}
  [\href{https://arxiv.org/abs/2008.06494}{{\ttfamily 2008.06494}}].

\bibitem{Klappert:2019emp}
J.~Klappert and F.~Lange, \emph{{Reconstructing rational functions with
  FireFly}}, \href{https://doi.org/10.1016/j.cpc.2019.106951}{\emph{Comput.
  Phys. Commun.} {\bfseries 247} (2020) 106951}
  [\href{https://arxiv.org/abs/1904.00009}{{\ttfamily 1904.00009}}].

\bibitem{Klappert:2020aqs}
J.~Klappert, S.Y.~Klein and F.~Lange, \emph{{Interpolation of dense and sparse
  rational functions and other improvements in FireFly}},
  \href{https://doi.org/10.1016/j.cpc.2021.107968}{\emph{Comput. Phys. Commun.}
  {\bfseries 264} (2021) 107968}
  [\href{https://arxiv.org/abs/2004.01463}{{\ttfamily 2004.01463}}].

\bibitem{Smirnov:2020quc}
A.V.~Smirnov and V.A.~Smirnov, \emph{{How to choose master integrals}},
  \href{https://doi.org/10.1016/j.nuclphysb.2020.115213}{\emph{Nucl. Phys. B}
  {\bfseries 960} (2020) 115213}
  [\href{https://arxiv.org/abs/2002.08042}{{\ttfamily 2002.08042}}].

\bibitem{Usovitsch:2020jrk}
J.~Usovitsch, \emph{{Factorization of denominators in integration-by-parts
  reductions}},  \href{https://arxiv.org/abs/2002.08173}{{\ttfamily
  2002.08173}}.

\bibitem{Kotikov:1990kg}
A.V.~Kotikov, \emph{{Differential equations method: New technique for massive
  Feynman diagrams calculation}},
  \href{https://doi.org/10.1016/0370-2693(91)90413-K}{\emph{Phys. Lett. B}
  {\bfseries 254} (1991) 158}.

\bibitem{Kotikov:1991pm}
A.V.~Kotikov, \emph{{Differential equation method: The Calculation of N point
  Feynman diagrams}},
  \href{https://doi.org/10.1016/0370-2693(91)90536-Y}{\emph{Phys. Lett. B}
  {\bfseries 267} (1991) 123}.

\bibitem{Kotikov:1991hm}
A.V.~Kotikov, \emph{{Differential equations method: The Calculation of vertex
  type Feynman diagrams}},
  \href{https://doi.org/10.1016/0370-2693(91)90834-D}{\emph{Phys. Lett. B}
  {\bfseries 259} (1991) 314}.

\bibitem{Remiddi:1997ny}
E.~Remiddi, \emph{{Differential equations for Feynman graph amplitudes}},
  \href{https://doi.org/10.1007/BF03185566}{\emph{Nuovo Cim. A} {\bfseries 110}
  (1997) 1435} [\href{https://arxiv.org/abs/hep-th/9711188}{{\ttfamily
  hep-th/9711188}}].

\bibitem{Gorishnii:1989dd}
S.G.~Gorishnii, \emph{{Construction of Operator Expansions and Effective
  Theories in the Ms Scheme}},
  \href{https://doi.org/10.1016/0550-3213(89)90622-6}{\emph{Nucl. Phys. B}
  {\bfseries 319} (1989) 633}.

\bibitem{Smirnov:1990rz}
V.A.~Smirnov, \emph{{Asymptotic expansions in limits of large momenta and
  masses}}, \href{https://doi.org/10.1007/BF02102092}{\emph{Commun. Math.
  Phys.} {\bfseries 134} (1990) 109}.

\bibitem{Smirnov:1994tg}
V.A.~Smirnov, \emph{{Asymptotic expansions in momenta and masses and
  calculation of Feynman diagrams}},
  \href{https://doi.org/10.1142/S0217732395001617}{\emph{Mod. Phys. Lett. A}
  {\bfseries 10} (1995) 1485}
  [\href{https://arxiv.org/abs/hep-th/9412063}{{\ttfamily hep-th/9412063}}].

\bibitem{Smirnov:2002pj}
V.A.~Smirnov, \emph{{Applied asymptotic expansions in momenta and masses}},
  {\emph{Springer Tracts Mod. Phys.} {\bfseries 177} (2002) 1}.

\bibitem{Schroder:2005va}
Y.~Schroder and A.~Vuorinen, \emph{{High-precision epsilon expansions of
  single-mass-scale four-loop vacuum bubbles}},
  \href{https://doi.org/10.1088/1126-6708/2005/06/051}{\emph{JHEP} {\bfseries
  06} (2005) 051} [\href{https://arxiv.org/abs/hep-ph/0503209}{{\ttfamily
  hep-ph/0503209}}].

\bibitem{Aglietti:2004tq}
U.~Aglietti and R.~Bonciani, \emph{{Master integrals with 2 and 3 massive
  propagators for the 2 loop electroweak form-factor - planar case}},
  \href{https://doi.org/10.1016/j.nuclphysb.2004.07.018}{\emph{Nucl. Phys. B}
  {\bfseries 698} (2004) 277}
  [\href{https://arxiv.org/abs/hep-ph/0401193}{{\ttfamily hep-ph/0401193}}].

\bibitem{vonManteuffel:2017hms}
A.~von Manteuffel and L.~Tancredi, \emph{{A non-planar two-loop three-point
  function beyond multiple polylogarithms}},
  \href{https://doi.org/10.1007/JHEP06(2017)127}{\emph{JHEP} {\bfseries 06}
  (2017) 127} [\href{https://arxiv.org/abs/1701.05905}{{\ttfamily
  1701.05905}}].

\bibitem{Beneke:1997zp}
M.~Beneke and V.A.~Smirnov, \emph{{Asymptotic expansion of Feynman integrals
  near threshold}},
  \href{https://doi.org/10.1016/S0550-3213(98)00138-2}{\emph{Nucl. Phys. B}
  {\bfseries 522} (1998) 321}
  [\href{https://arxiv.org/abs/hep-ph/9711391}{{\ttfamily hep-ph/9711391}}].

\bibitem{Smirnov:1998vk}
V.A.~Smirnov and E.R.~Rakhmetov, \emph{{The Strategy of regions for asymptotic
  expansion of two loop vertex Feynman diagrams}},
  \href{https://doi.org/10.1007/BF02557396}{\emph{Theor. Math. Phys.}
  {\bfseries 120} (1999) 870}
  [\href{https://arxiv.org/abs/hep-ph/9812529}{{\ttfamily hep-ph/9812529}}].

\bibitem{Smirnov:1999bza}
V.A.~Smirnov, \emph{{Problems of the strategy of regions}},
  \href{https://doi.org/10.1016/S0370-2693(99)01061-8}{\emph{Phys. Lett. B}
  {\bfseries 465} (1999) 226}
  [\href{https://arxiv.org/abs/hep-ph/9907471}{{\ttfamily hep-ph/9907471}}].

\bibitem{Pak:2010pt}
A.~Pak and A.~Smirnov, \emph{{Geometric approach to asymptotic expansion of
  Feynman integrals}},
  \href{https://doi.org/10.1140/epjc/s10052-011-1626-1}{\emph{Eur. Phys. J. C}
  {\bfseries 71} (2011) 1626}
  [\href{https://arxiv.org/abs/1011.4863}{{\ttfamily 1011.4863}}].

\bibitem{Jantzen:2012mw}
B.~Jantzen, A.V.~Smirnov and V.A.~Smirnov, \emph{{Expansion by regions:
  revealing potential and Glauber regions automatically}},
  \href{https://doi.org/10.1140/epjc/s10052-012-2139-2}{\emph{Eur. Phys. J. C}
  {\bfseries 72} (2012) 2139}
  [\href{https://arxiv.org/abs/1206.0546}{{\ttfamily 1206.0546}}].

\bibitem{Panzer:2014caa}
E.~Panzer, \emph{{Algorithms for the symbolic integration of hyperlogarithms
  with applications to Feynman integrals}},
  \href{https://doi.org/10.1016/j.cpc.2014.10.019}{\emph{Comput. Phys. Commun.}
  {\bfseries 188} (2015) 148}
  [\href{https://arxiv.org/abs/1403.3385}{{\ttfamily 1403.3385}}].

\bibitem{Czakon:2005rk}
M.~Czakon, \emph{{Automatized analytic continuation of Mellin-Barnes
  integrals}}, \href{https://doi.org/10.1016/j.cpc.2006.07.002}{\emph{Comput.
  Phys. Commun.} {\bfseries 175} (2006) 559}
  [\href{https://arxiv.org/abs/hep-ph/0511200}{{\ttfamily hep-ph/0511200}}].

\bibitem{PSLQ}
H.~Ferguson and D.~Bailey, \emph{A polynomial time, numerically stable integer
  relation algorithm}, {\emph{RNR Technical Report} (1992) }.

\bibitem{Liu:2017jxz}
X.~Liu, Y.-Q.~Ma and C.-Y.~Wang, \emph{{A Systematic and Efficient Method to
  Compute Multi-loop Master Integrals}},
  \href{https://doi.org/10.1016/j.physletb.2018.02.026}{\emph{Phys. Lett. B}
  {\bfseries 779} (2018) 353}
  [\href{https://arxiv.org/abs/1711.09572}{{\ttfamily 1711.09572}}].

\bibitem{Liu:2022mfb}
Z.-F.~Liu and Y.-Q.~Ma, \emph{{Determining Feynman Integrals with Only Input
  from Linear Algebra}},
  \href{https://doi.org/10.1103/PhysRevLett.129.222001}{\emph{Phys. Rev. Lett.}
  {\bfseries 129} (2022) 222001}
  [\href{https://arxiv.org/abs/2201.11637}{{\ttfamily 2201.11637}}].

\bibitem{Liu:2022chg}
X.~Liu and Y.-Q.~Ma, \emph{{AMFlow: A Mathematica package for Feynman integrals
  computation via auxiliary mass flow}},
  \href{https://doi.org/10.1016/j.cpc.2022.108565}{\emph{Comput. Phys. Commun.}
  {\bfseries 283} (2023) 108565}
  [\href{https://arxiv.org/abs/2201.11669}{{\ttfamily 2201.11669}}].

\bibitem{Kinoshita:1962ur}
T.~Kinoshita, \emph{{Mass singularities of Feynman amplitudes}},
  \href{https://doi.org/10.1063/1.1724268}{\emph{J. Math. Phys.} {\bfseries 3}
  (1962) 650}.

\bibitem{Lee:1964is}
T.D.~Lee and M.~Nauenberg, \emph{{Degenerate Systems and Mass Singularities}},
  \href{https://doi.org/10.1103/PhysRev.133.B1549}{\emph{Phys. Rev.} {\bfseries
  133} (1964) B1549}.

\bibitem{Harlander:2020cyh}
R.V.~Harlander, S.Y.~Klein and M.~Lipp, \emph{{FeynGame}},
  \href{https://doi.org/10.1016/j.cpc.2020.107465}{\emph{Comput. Phys. Commun.}
  {\bfseries 256} (2020) 107465}
  [\href{https://arxiv.org/abs/2003.00896}{{\ttfamily 2003.00896}}].

\bibitem{mathfile}
  \emph{ggHNiggetiedtUsovitsch.m}, attached.

\end{thebibliography}\endgroup

\end{document}